# Tractions and stress fibers control cell shape and rearrangements in collective cell migration

Aashrith Saraswathibhatla, Jacob Notbohm*

Department of Engineering Physics

University of Wisconsin-Madison

1500 Engineering Dr

Madison, WI, 53706

* Corresponding author:

Jacob Notbohm

jknotbohm@wisc.edu

+1.608.890.0030

1500 Engineering Dr

Madison, WI, 53706

**Abstract**

Key to collective cell migration is the ability of cells to rearrange their position with respect to their neighbors. Recent theory and experiments demonstrated that cellular rearrangements are facilitated by cell shape, with cells having more elongated shapes and greater perimeters more easily sliding past their neighbors within the cell layer. Though it is thought that cell perimeter is controlled primarily by cortical tension and adhesion at each cell's periphery, experimental testing of this hypothesis has produced conflicting results. Here we studied collective migration in an epithelial monolayer by measuring forces, cell perimeters, and motion, and found all three to decrease with either increased cell density or inhibition of cell contraction. In contrast to previous understanding, the data suggest that cell shape and rearrangements are controlled not by cortical tension or adhesion at the cell periphery but rather by the stress fibers that produce tractions at the cell-substrate interface. This finding is confirmed by an experiment showing that increasing tractions reverses the effect of density on cell shape and rearrangements. Our study therefore reduces the focus on the cell periphery by establishing cell-substrate traction as a major physical factor controlling shape and motion in collective cell migration.

## I. INTRODUCTION

In numerous cases in human health and disease, epithelial cells transition from a static, motionless state to an active, migratory state. The active cell migration may generate new tissue, as in embryonic development, or further the spread of disease, as in cancer progression [1–3]. In either case, a critical step is the transition from motionless to migratory. Experiments have observed this transition [3–8], and demonstrated that it is akin to the jamming transition that occurs in particulate matter [9,10]. By analogy to the effect of density on jamming of rigid particles [9,10], one might expect the transition from migratory to motionless in a cell collective to result from an increase in cell number density, and indeed this prediction has been observed in experiments [4,5,7,11–14]. How the cell density affects the collective cell migration remains unclear, however, because the analogy between granular materials and cells is imperfect. In granular materials, the density-induced jamming transition is related to packing of particles and the free space available for motion, but cells within a monolayer cover all free space [15].

The mechanism causing jamming must therefore differ for cell monolayers and granular materials, though the differences remain unclear.

It may be that the effects of density on motion are indirect, which would mean that the motion is controlled by some other underlying physical factor. In support of this, experiments have observed greater phosphorylated myosin light chain in epithelial monolayers of low density compared to high density [14], which is consistent with experiments that observed a proportional relationship between cell size and contractile stress [16,17]. This implies that density affects the active forces produced by each cell. Additionally, an increase in cell density due to cell division in epithelial monolayers coincides with maturation of cadherin-based cell-cell adhesions, which was proposed to reduce migration by increasing an effective friction between neighboring cells [7]. Thus, density may also affect adhesion between cells. However, there remains no clear link between density, adhesion, force, and motion.

A useful starting point to relate these various factors is a set of theoretical models relating the energy of each cell $E$ to the cell's area $A$ and perimeter $P$ according to the equation $E = K_A(A - A_0)^2 + \xi P^2 + \Gamma P$, where $K_A, A_0, \xi,$ and $\Gamma$ are constants [15,18–26]. The variable $\Gamma$ is a line tension that acts much like surface tension in a fluid. It is thought to be controlled by the mechanical components at each cell-cell interface. More specifically, tension within the cell cortex increases the line tension, whereas cell-cell adhesion molecules decrease it [6,15,18–27]. The theory predicts that a parameter defining the average cell shape, $q = P/\sqrt{A}$, is a metric to determine whether the cell layer is solid-like, with no rearrangements between neighboring cells, or fluid-like, with cells easily rearranging and sliding past their neighbors [6,21,28]. As $q$ is linearly proportional to the cell perimeter, we refer to it hereafter as the cell perimeter.

Though experiments have confirmed the relationship between perimeter and migration [6], the interpretation of the theory is still unclear. Firstly, as the cell perimeter is dimensionless, it is independent of cell size, and thus it is unclear whether it is affected by density. Secondly, some implementations of the theory assume the line tension $\Gamma$ to be negative [6,15,20,21,23–26], implying that the effect of cell-cell adhesions on the line tension is greater in magnitude than the effect of cortical tension. Experiments have not yet verified this assumption. Thirdly, some experiments relating perimeter to motion in monolayers of cells also measured cell-substrate tractions and observed greater traction applied by fast-moving cells with larger perimeter [6,28]. The theory predicts that increased actomyosin contraction would increase cortical tension thereby reducing average cell perimeter and tending to diminish collective rearrangements. A possible resolution for this inconsistency is that traction itself may increase the cell perimeter and facilitate rearrangements. This has been suggested by a theoretical model that introduced a "self-propulsion force," which is equivalent to traction [21], but this theoretical prediction has not been tested.

Here we compare the effects of line tension at the cell periphery and traction at the cell-substrate interface on perimeter and rearrangements in monolayers of epithelial cells. In response to increased number density or decreased actomyosin contractility, cell traction, perimeter, and rearrangements decreased. We further show that by activating actomyosin contraction, the established relationship between density and rearrangement can be reversed. The commonality underlying all experiments is the relationship between tractions, cell perimeter, and rearrangements, in agreement with a theoretical prediction of the effect of traction on perimeter and rearrangements [21]. Our results therefore point to traction as a major factor controlling both shape and motion in collective cell migration.

## II. RESULTS

We began with the well-established observation that cell density affects collective cell migration, with higher density tending to arrest rearrangements within a cell monolayer [4,5,7,11–14] . We seeded Madin-Darby canine kidney (MDCK) type II cells into 1 mm micropatterned islands on polyacrylamide substrates with Young's modulus of 6 kPa coated with collagen I. Cells were seeded to attain densities of approximately 1200 and 4200 cells/mm$^2$ by adjusting the seeding density but keeping the time of culture constant (Fig. 1a, c). Cells were imaged over time, and cell trajectories showed a difference in motion, with cells in high density islands remaining stuck in place but cells in low density islands moving more freely (Fig. 1b, d; Fig. S1a-b). Notably, in low density islands, collective swirls can be observed (Fig. S1a), indicating the presence of collective packs. For one pack to slide past another, cells must rearrange their local position with their neighbors. At the spatial resolution of our experiments, cell rearrangements could not be tracked directly, so we inferred cell rearrangements from two metrics of collective migration. First, we computed the mean square displacement (MSD, see Eq. 1a of Methods) of all cell trajectories in each island over a time interval of 120 minutes (Fig. 1e). The MSD grows quadratically in time if cells move in a straight line; it grows more slowly than this if cell trajectories bend or curve, as would occur if cell motion is impeded by neighboring cells. We therefore fit the MSD to a function of the form MSD ~ $\Delta t^\alpha$ with the exponent $\alpha$ quantifying the cell motion. Consistent with previous studies [4,5,7,11–14], the MSD for all cells in high density islands had a smaller value of $\alpha$ than cells in low density islands (Fig. 1f), indicating more constrained migration. We also analyzed the MSD of individual cell trajectories averaging only over time (MSD$_i$, Fig. S1c). The corresponding exponent $\alpha_i$ was significantly larger in low density islands (Fig. S1d), which provides additional evidence that more rearrangements are present in low density islands compared to high density islands. However, two different mechanisms could produce a large exponent $α$: (1) cells could freely rearrange with their neighbors or (2) a collective group of cells could move in a straight line. To verify that the differences in $α$ resulted from differences in rearrangements rather than straight line motion of collective groups, we used a second metric, the cage-relative MSD (CRMSD, Eq. 1b of Methods), which analyzes the difference between each cell's displacement and the displacement of its nearest $N$ neighbors, thereby quantifying the local rearrangement between that cell and its neighbors [29]. The CRMSD was fit to CRMSD $\sim \Delta t^{\alpha_{CR}}$ with the exponent $α_{CR}$ quantifying the cage-relative cell motion. As previous modeling and experiments showed cells to move in packs of size ~20 cells [4,29], we varied $N$ from 4 to 20 cells and observed no significant change in $α_{CR}$ (Fig. S1e). We therefore chose a pack size of $N = 20$ cells for the remainder of our analysis. Compared to $α$, $α_{CR}$ is smaller, but the trends are nonetheless the same, namely that $α_{CR}$ is smaller in islands of higher density (Fig. 1g, S1f). As both metrics for rearrangements, $α$ and $α_{CR}$, are decreased by increasing density, the data suggests increasing density reduces rearrangements, as in previous studies [4,5,7,11–14].

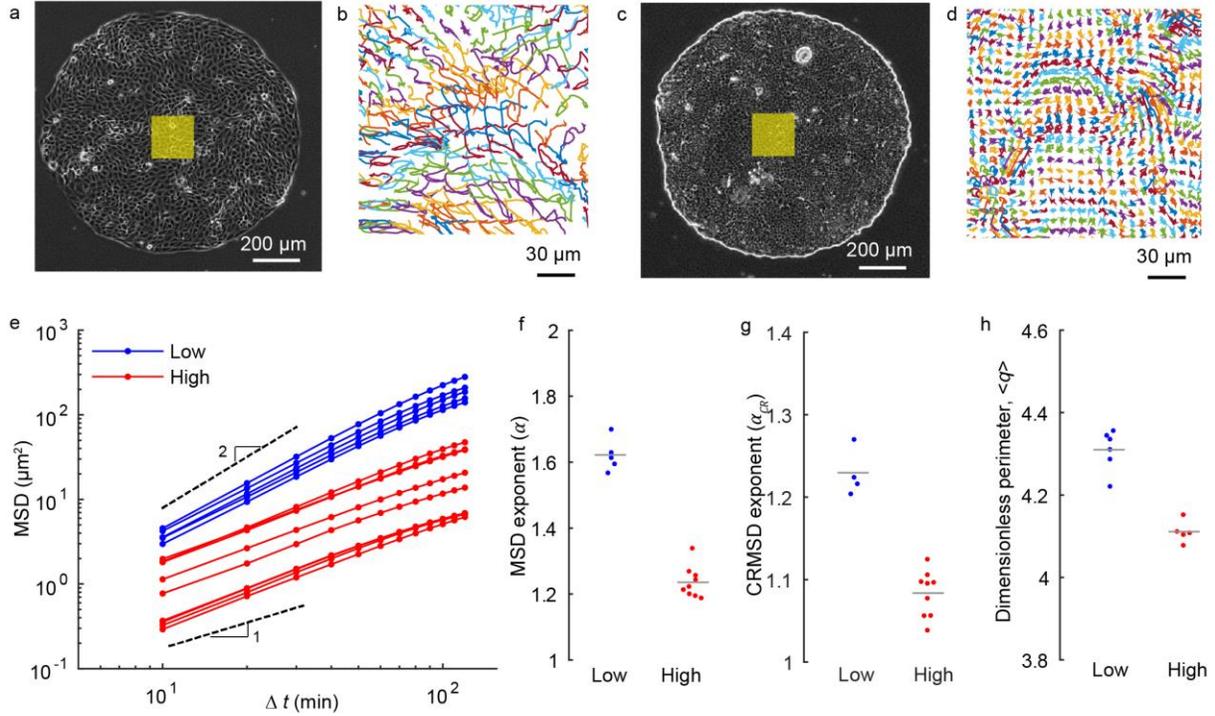

FIG. 1. Cell density decreases cell rearrangements and perimeter $q$. (a, c) Phase contrast images of low density (1200 cells/mm$^2$) (a) and high density (4200 cells/mm$^2$) (c) cell islands. (b, d) Cell trajectories for the highlighted regions over 8 hours in islands of low (b) and high (d) density. Trajectories for the entire islands are shown in Fig. S1. (e) Mean square displacement (MSD) for low and high density islands plotted on logarithmic axes. In this and all subsequent figures, each line of MSD corresponds to a different cell island; dots represent the individual data points that make up the curve. (f) The MSD is fit to MSD ~ $t^\alpha$, with exponent $\alpha$ characterizing cell rearrangements. High density islands have a smaller value of $\alpha$ than low density islands ($p < 0.001$). In this and all subsequent figures, each dot of $\alpha$ corresponds to a different cell island. (g) Low density islands have a larger exponent $\alpha_{CR}$ than high density islands ($p < 0.001$). (h) Average perimeter $<q>$ is greater for cells at low density compared to high density ($p < 0.001$). In this and all subsequent figures, each dot of $<q>$ corresponds to an average over 100-150 cells in a field of view, with at least 2 fields of view from at least 3 biologically different samples.

Consistent with previous experiments [6,28], the reduction in rearrangements also coincided with a decreased average perimeter $<q>$ (see Methods for definition, Fig. 1h). As in previous studies [8], the average cell aspect ratio showed the same trends as perimeter, namely that cells were more elongated in islands of low density than high (Fig. S2). For the remainder of this manuscript, we focus on the perimeter $q$, as it is the variable considered by the theoretical models.

Because $q$ is dimensionless, it is unclear how it is affected by cell density, which has units of inverse area. For $q$ to change, the perimeter $P$ (having units of length) must scale non-proportionally with the square root of area $A^{1/2}$. We hypothesized the non-proportional scaling to result from a change in the distribution of forces controlling cell perimeter and area. We began by considering the forces at the cell periphery. According to current understanding [6,15,18–27], a decrease in the perimeter $q$ is caused by an increase in the line tension $\Gamma$, which can result from an increase in cortical tension or a decrease in cell-cell adhesion.

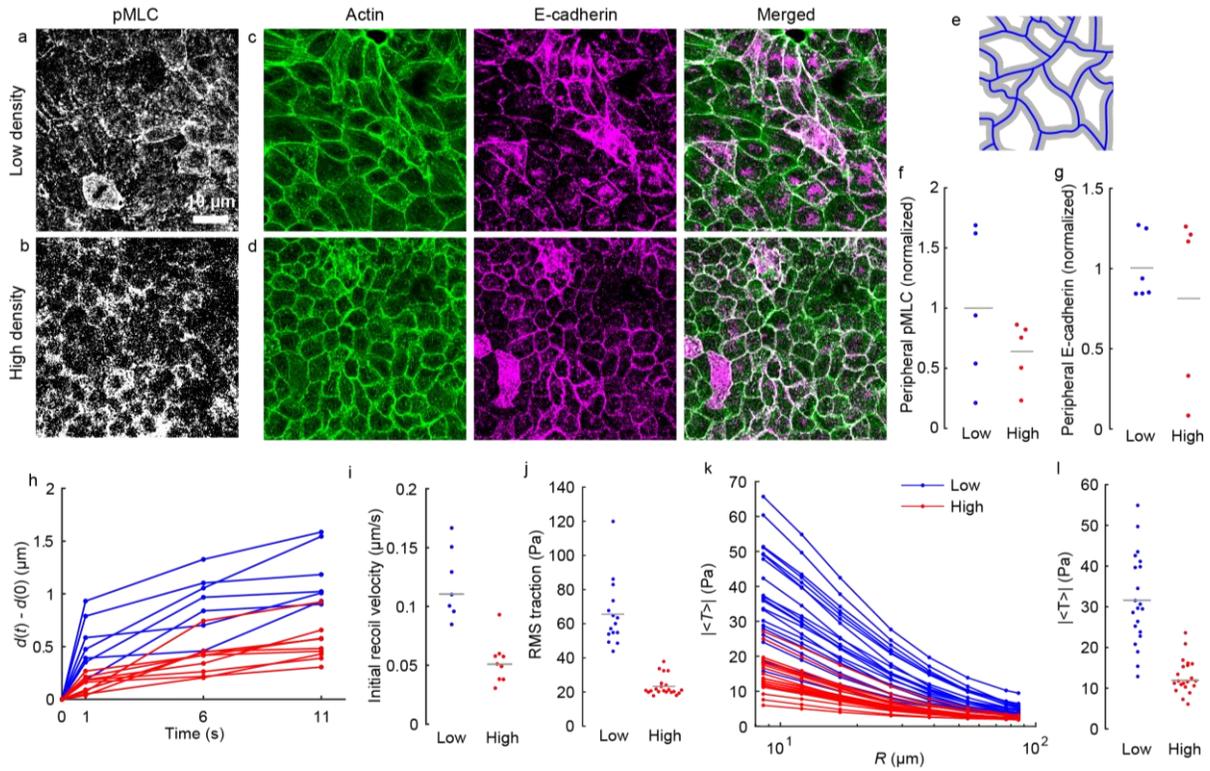

FIG. 2. Cell density has no effect on cortical tension or E-cadherin but decreases tractions. (a, b) Confocal images of phosphorylated myosin light chain (pMLC) at low (1200 cells/mm$^2$) and high density (4200 cells/mm$^2$). (c, d) Confocal images of actin and E-cadherin for cells at low (c) and high (d) density. (e) Drawing showing analysis of fluorescent intensities at cell peripheries. Segmented cell boundaries (blue) are dilated by 1 μm (gray); pixels within the dilated region are used for quantifying peripheral fluorescent intensities. (f) Fluorescent intensity of peripheral pMLC was unaffected by cell density ($p = 0.4$). For all data quantifying fluorescent images in this manuscript, each dot corresponds to an average over 100-150 cells in a field of view, with at least 2 fields of view from at least 3 biologically different samples. (g) Fluorescent intensity of peripheral E-cadherin was unaffected by cell density ($p = 0.7$). (h) Change in distance between the vertices of an ablated edge, $d(t) − d(0)$, in cells at low and high densities. Each line corresponds to an ablated cell edge. (i) Cells at low density show higher initial recoil velocity compared to cells at high density ($p < 0.01$). For all data quantifying recoil velocity in this manuscript, each dot corresponds to a different ablated edge. (j) Average RMS traction decreases with increase in cell density ($p < 0.001$). (k) Average traction imbalance computed over different circular regions of radius $R$. In this and all subsequent figures quantifying traction or traction imbalance, each dot corresponds to a different cell island. (l) Average traction imbalance for $R = 12$ μm is statistically larger in islands of low density than high density ($p < 0.001$).

To assess cortical tension, we fluorescently stained for phosphorylated myosin light chain (pMLC, Fig. 2a-b), indicative of actomyosin contractility which generates cortical tension [27]. Using fluorescent images of actin (Fig. 2c-d) to detect cell peripheries (Fig. 2e and Methods), we quantified fluorescent intensities of peripheral pMLC in cells at low and high density. Surprisingly, density had no apparent effect on peripheral pMLC intensity, suggesting no change in cortical tension (Fig. 2a-b, f). For a second measure of cortical tension, we performed laser ablation of cell-cell edges. The recoil of vertices previously connecting an ablated edge correlates positively to the tension in that edge before ablation and negatively to friction at that edge [30]. Consistent with this understanding, peripheral pMLC fluorescence and recoil velocity from laser ablation correlates in epithelial systems such as Caco-2 cells and Drosophila [31,32]. We ablated cell edges (Supplemental Video 1) and measured the distance $d(t)$ between vertices that connected the ablated edge from 1 to 41 s after ablation. Data were reported by taking the change in distance, $d(t) - d(0)$, where $d(0)$ was defined as the initial length of the cell edge before ablation. Most vertices retracted linearly with time for 11 s after which they retracted more slowly (Fig. S3). We therefore fit a line to the change in distance for $t = 0$ to 11 s (gray region in Fig. S3) to compute the initial recoil velocity. Cells at low density showed higher recoil velocity compared to cells at high density (Fig. 2h-i). As there was no change in pMLC fluorescent intensity, this finding may suggest greater friction at cell-cell edges in islands of greater density as proposed previously [7]. It is also possible that laser ablation is more sensitive than pMLC fluorescence to line tension, implying that increasing density causes a decrease in cortical tension, in agreement with previous findings in epidermal progenitor cells [35].

We next considered cell-cell adhesion by fluorescent intensity of E-cadherin, which is associated with changes in cell-cell contact areas and cell sorting [36,37]. Fluorescent imaging showed no difference in E-cadherin at low and high densities implying no change in cell-cell adhesions (Fig. 2c, d, g). As the results suggest that increasing density had no effect on adhesion and caused either no change or a reduction in cortical tension, we conclude that increased density either had no effect on line tension $\Gamma$ or decreased it. Accordingly, the average perimeter $<q>$ would be expected to either remain the same or increase, but in contrast, increasing density decreased $<q>$ (Fig. 1h).

We then quantified traction at the cell-substrate interface, which has been proposed by a theoretical model to be another factor that increases the average perimeter $<q>$ [21]. We measured tractions in the islands of different cell density using traction force microscopy (Fig. S4a-b) [38–40]. RMS tractions were larger in islands of low density (Fig. 2j), consistent with the inverse relationship between cell density and cell contraction observed in other studies [11,14,16,17]. As the RMS gives an average over space, it does not necessarily quantify the traction that produces motion. It could be that the tractions produced by each cell balance such that the vector sum of traction applied by that cell to the substrate is zero. To account for this possibility, we defined a traction imbalance (see Methods), which quantifies the vector sum of traction for different circular regions of radius $R$. A nonzero traction imbalance implies stresses at the cell-cell interfaces [11,41], and is a driving force for motion. The average traction imbalance was larger in low density islands than high (Fig. 2k), which is consistent with the trend in RMS traction. A statistical test on the traction imbalance for $R = 12$ μm showed a significant difference (Fig. 2l). These findings hint that traction at the cell-substrate interface, rather than cortical tension or adhesion at the cell periphery, may be the physical factor controlling cell shape and motion.

To investigate further the effects of adhesion, cortical tension, and traction on shape and motion, we designed experiments to modulate actomyosin contraction by treating with blebbistatin to disrupt myosin II and cytochalasin D to disrupt actin polymerization in islands of equal density (1200 cells/mm$^2$). To investigate the effect of the inhibitors on cortical tension, we imaged the actin and pMLC at the cell peripheries (Fig. 3a, c). Blebbistatin caused cell peripheries to become more curved (Fig. 3a-b) implying reduced cortical tension [28,31], whereas cytochalasin D decreased the amount of peripheral pMLC (Fig. 3c-d), also implying reduced cortical tension [31,32]. Additionally, recoil velocity decreased in response to either treatment (Supplemental Video 1, Fig. 3e-h), which, assuming no change in cell-cell friction, indicates reduced cortical tension. These observations are consistent with the fact that cortical tension results from actomyosin contractility [27]. To assess cell-cell adhesions, we fluorescently labeled E-cadherin. The treatments had no significant effect on E-cadherin intensities (Fig. S5), indicating no effect of the treatments on adhesion. Together, these findings suggest that actomyosin inhibition reduces the line tension $\Gamma$. According to current understanding [6,15,18–27], the reduced line tension should increase the average cell perimeters and the rearrangements. By contrast, the treatments reduced the cell perimeters (Fig. 3i-j). Additionally, both inhibitors reduced the exponents $\alpha$ of MSD and $\alpha_{CR}$ of CRMSD (Fig. S6), implying reduced cell rearrangements. These results contrast with current understanding, that decreased line tension should increase cell perimeter and rearrangements, which suggests some other factor likely affects perimeter and rearrangements.

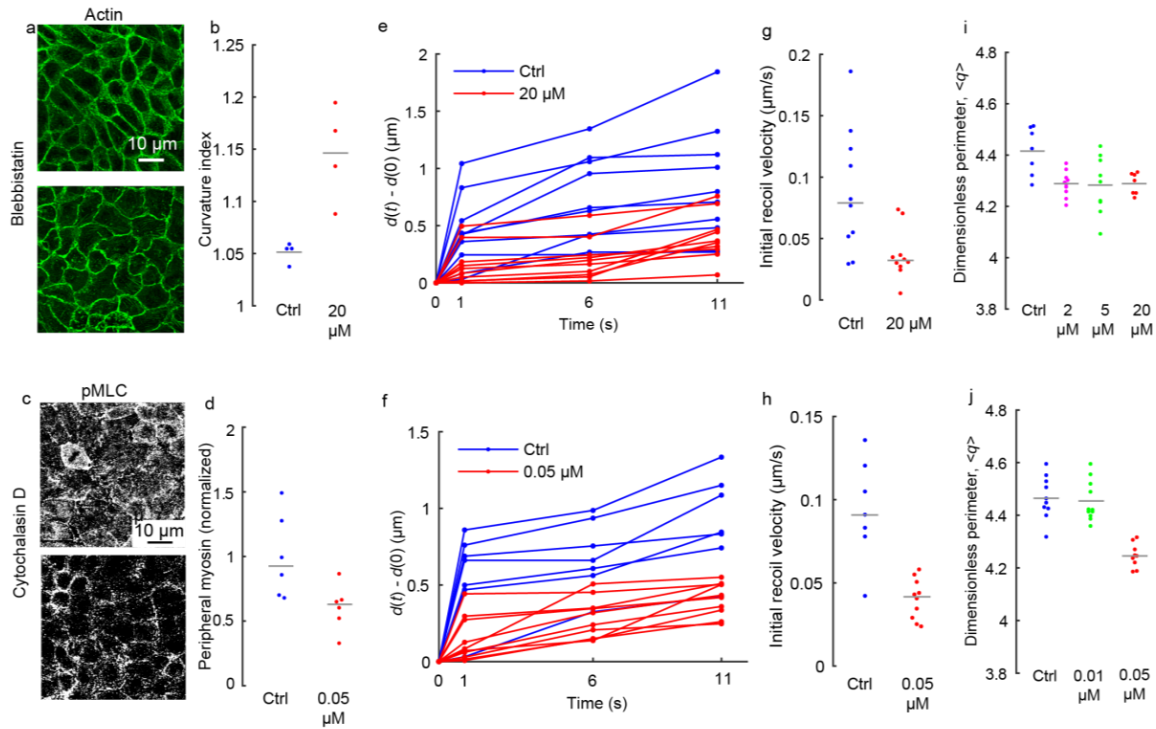

FIG. 3. Effect of blebbistatin and cytochalasin D on cortical tension and cell perimeter. (a) Cortical actin imaged for control cells (top) or cells treated with 20 μM blebbistatin (bottom). (b) Cell curvature index (defined as total edge length divided by end-to-end distance) increases after treating with 20 μM blebbistatin ($p < 0.01$). (c) Phosphorylated myosin light chain (pMLC) imaged for control cells (top) or cells treated with 0.05 μM cytochalasin D (bottom). (d) Peripheral pMLC fluorescent intensity is decreased by cytochalasin D ($p < 0.05$). (e, f) Change in distance between the vertices of an ablated edge, $d(t) - d(0)$, for control and cells treated with 20 μM blebbistatin (e) or 0.05 μM cytochalasin D (f). (g, h) Initial recoil velocity decreased after treating with 20 μM blebbistatin ($p < 0.01$, g) or 0.05 μM cytochalasin D ($p < 0.001$, h). (i, j) Average perimeter $<q>$ measured after treating with blebbistatin (i) or cytochalasin D (j) at different concentrations. Both 20 μM blebbistatin ($p < 0.01$) and 0.05 μM cytochalasin D ($p < 0.001$) decrease $<q>$. For all data in this figure, the cell density is approximately 1200 cells/mm$^2$.

Tractions were also measured before and after inhibiting contractility with blebbistatin or cytochalasin D. In these experiments, the RMS traction was measured before and after each treatment and normalized to the value before the treatment. The RMS tractions were reduced by both treatments in a dose-dependent manner and remained nearly constant for 1 to 3 hours after each treatment (Fig. 4a, b, Fig. S4c, d). Averaging over this time window gave a single scalar measure of traction for each cell island. Statistical comparison showed that both inhibitors significantly decreased the RMS traction (Fig. 4c, d). Likewise, the traction imbalance was significantly reduced by the inhibitors (Fig. 4e-h). The data therefore suggest that traction, rather than cortical tension or cell-cell adhesion, may affect cell perimeters and rearrangements. This is in contrast with the notion that perimeter and rearrangements result from cortical tension and cell-cell adhesion [6,15,18–27], but it is nevertheless consistent with a theoretical model that predicted traction, in addition to cortical tension and adhesion, can increase cell perimeter and rearrangements [21].

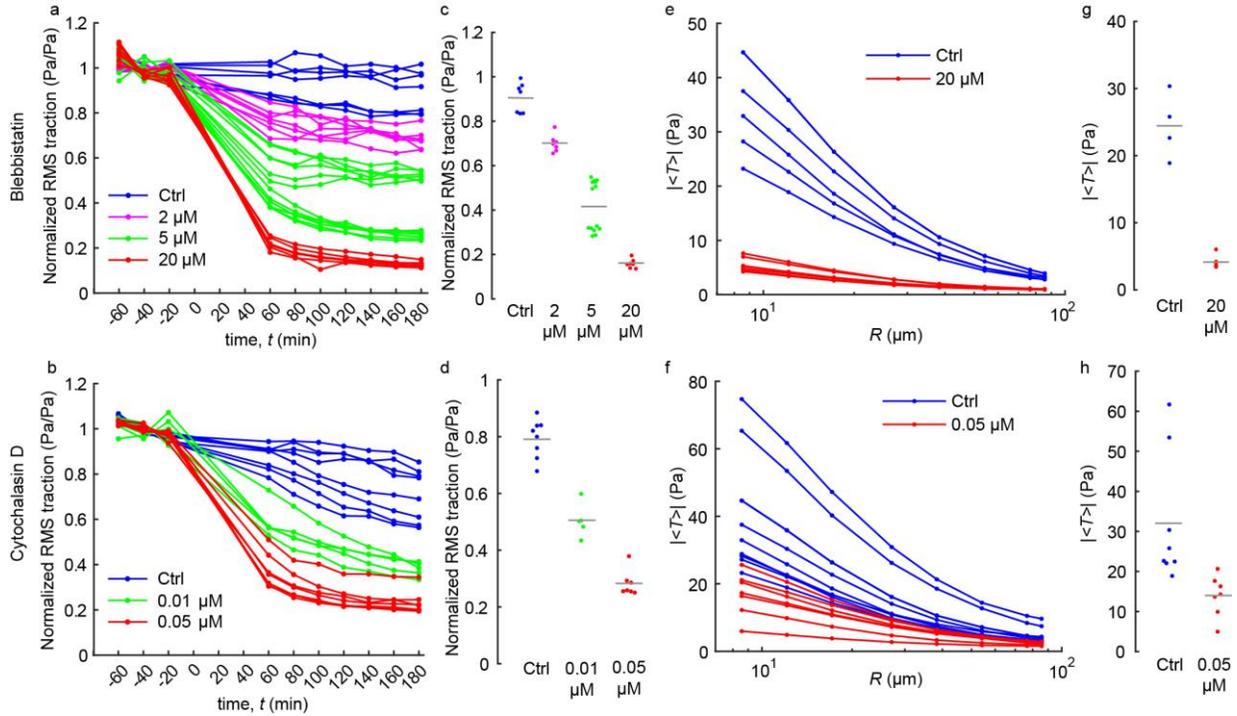

FIG. 4. Blebbistatin and cytochalasin D show dose-dependent reduction in tractions. (a, b) Average RMS traction for different cell islands measured over time before ($t < 0$) and after ($t > 0$) treating with blebbistatin (a) or cytochalasin D (b). For each island, the average RMS traction is normalized by its average value before the treatment. (c, d) Normalized RMS tractions are averaged for $t = 60$–$120$ min. All groups are statistically different from each other ($p < 0.001$). (e, f) Average traction imbalance over different circular regions of radius $R$ for control islands and islands treated with blebbistatin (e), and cytochalasin D (f). (g, h) The magnitude of average traction imbalance for $R = 12$ μm is reduced by both blebbistatin (g) and cytochalasin D (h) ($p < 0.001$ for both treatments compared to control). For all data in this figure, the cell density is approximately 1200 cells/mm$^2$.

The notion that traction relates to the perimeter more so than cortical tension or adhesion is unexpected, because the forces of cell traction are not applied around the cell's periphery, as is cortical tension, but rather at the cell-substrate interface. The only logical explanation for tractions having a dominant effect on the perimeter is that the forces associated with traction must be far larger than forces of cortical tension and adhesion. This can be understood by the fact that the amount of pMLC in the stress fibers is approximately 3.5 times that in the cortex (Fig. S7). To explore this idea further, we imaged stress fibers for cells in monolayers of high (~4200 cells/mm$^2$) and low (~1200 cells/mm$^2$) density. The images showed a clear difference, with cells at high density having fewer and less aligned stress fibers (Fig. 5a, b), in agreement with the lower tractions produced (Fig. 2i). To quantify stress fiber alignment, we computed an order parameter $S$ (Eq. 2 of Methods) which attains a value of 0 when stress fibers are randomly aligned and 1 when they are fully aligned. The average value of order parameter $S$ was 0.59 for cells at low density and 0.43 for cells at high density (Fig. 5c). Similarly, both blebbistatin and cytochalasin D dramatically reduced the number of stress fibers (Fig. S5a-c).

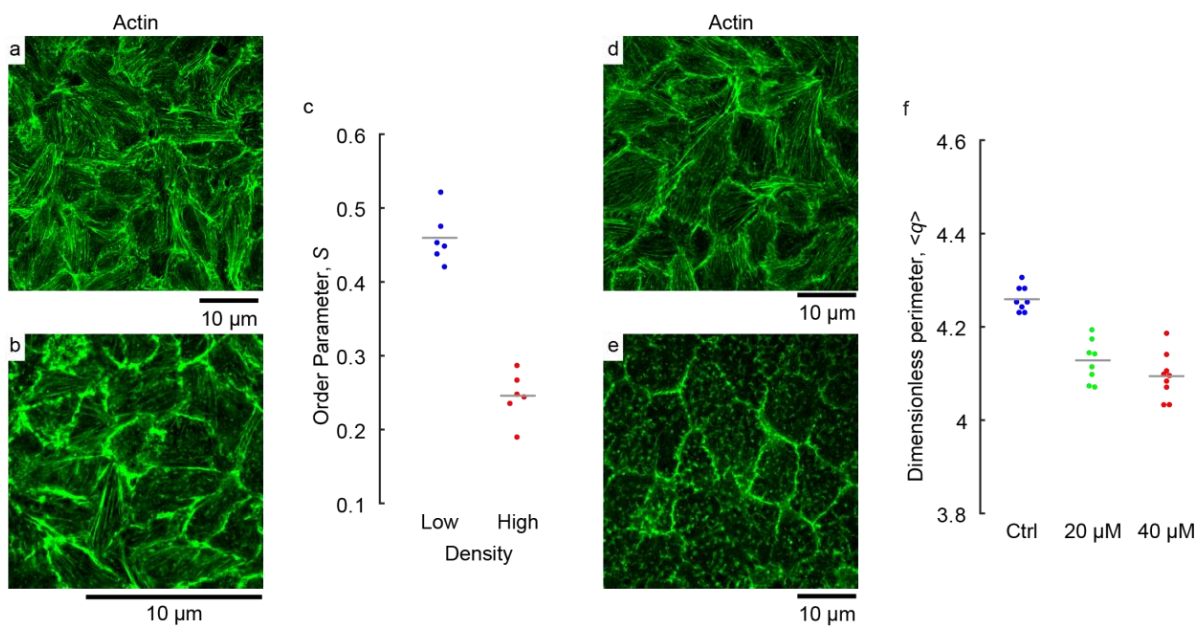

FIG. 5. Effect of density on stress fiber organization. (a, b) Confocal images of actin stress fibers in low (1200 cells/mm$^2$) (a) and high (4200 cells/mm$^2$) (b) density monolayers. (c) Order parameter $S$ quantifying average stress fiber alignment in different monolayers. $S$ is lower at high density compared to low density ($p < 0.001$). (d, e) Confocal images of actin stress fibers of control cells (d) and cells treated with 40 μM SMIFH2 (e) (the density for both panels is approximately 1200 cells/mm$^2$). (f) The average dimensionless perimeter $<q>$ decreases with SMIFH2 treatment ($p < 0.001$ for both treatments compared to control).

As a test of the relationship between stress fibers and perimeter, we used SMIFH2 to inhibit formins, which bundle actin filaments to form actin stress fibers [42]. Two concentrations (20 μM and 40 μM) of SMIFH2 decreased the traction by 60% and 65% respectively (Fig. S8a-b). The treatment reduced stress fibers and their alignment (Fig. 5d, e) as confirmed by the order parameter *S* (Fig. S8c). The effect of SMIFH2 on cell rearrangements was not quantified, because the effects of SMIFH2 wear off after ~1 hour of treatment [43]. Consistent with this timing, the tractions began to increase 1 hour after the treatment (Fig. S8a). Nevertheless, the inhibitor decreased the dimensionless perimeter <*q*> (Fig. 5f), giving additional evidence of the connection between stress fibers and average perimeter <*q*>. To rule out the effects of SMIFH2 on cell perimeter from altered cortical tension or adhesions, we investigated cortical tension by performing laser ablation on cell edges and staining for pMLC. In agreement with previous findings [44], initial recoil velocity was lower in cells with SMIFH2 compared to control, suggesting a reduction in cortical tension (Fig. S8d-e). Surprisingly, pMLC fluorescent intensity at the cell peripheries increased (Fig. S8f), implying an increase in cortical tension. We suspect that the difference in observations between recoil velocity and pMLC is that formin inhibition can lead to disorganized F-actin at cell-cell junctions [44]. Therefore, even though there may be more pMLC at cell peripheries, the myosin may be disorganized and unrelated to cortical tension. Consistent with this explanation, laser ablation of SMIFH2-treated cell edges caused cortical F-actin to disperse rather than retract (Supplemental Video 1), suggesting disorganized cortical actomyosin. The SMIFH2 treatment also decreased E-cadherin expression at cell-cell junctions (Fig. S8g), consistent with the notion that formin-nucleated actin at cell-cell junctions stabilizes cell-cell adhesions [44]. In summary, the effect of formin inhibition on line tension remains unclear, with pMLC fluorescence and E-cadherin imaging suggesting an increase in line tension and laser ablation suggesting a decrease. Nevertheless, the experiments with the formin inhibitor show a clear relationship between stress fibers (Fig. 5d-e, S8c) and average perimeter <*q*> (Fig. 5f), giving additional evidence of the connection between stress fibers and perimeter. Notably, when stress fibers were present, they spanned the entire cell-substrate interface, which had an area far greater than that of the cell-cell interface. It is therefore reasonable to infer that the forces of cell-substrate traction are far larger than forces in the cortex at cell-cell adhesions. This would explain how tractions affect the average cell perimeter even though they act inside the cell's boundaries rather than at its periphery.

In all our experiments, the data suggest that tractions produced by stress fibers are the primary driver behind cell perimeter and rearrangements. We therefore hypothesized that the effect of density on cell perimeter (Fig. 1h) and rearrangements (Fig. 1e-g) could be reversed by increasing cell-substrate tractions. To test this hypothesis, we seeded islands at two different densities, 1200 cells/mm$^2$ and 2600 cells/mm$^2$. As in Fig. 1, cells at higher density had a lower perimeter $<q>$ and rearranged less than cells at low density (Fig. 6a-d). To increase cell-substrate tractions in the higher density islands, we treated them with the Rho activator CN03 (2 μg/mL), which increased stress fiber alignment (Fig. S9a-d) and increased both the RMS traction and the traction imbalance by a factor of approximately 3 (Fig. 6e-f, Fig. S9e, f). The treatment also increased the perimeter $<q>$ such that it became statistically indistinguishable from the perimeter of cells at low density (Fig. 6a). As Figs. 3 and 4 suggest that traction affects cell perimeter, we expect the same to be true here. An alternative explanation is that CN03 treatment affected cell line tension by decreasing cortical tension, increasing cell-cell adhesions, or both. To rule out this explanation, we quantified fluorescence of pMLC and E-cadherin at cell peripheries and measured recoil velocity after laser ablation. There was no significant change in fluorescent intensities or recoil velocities (Fig. S10), implying no change in line tension. Another alternative explanation is that the CN03 treatment directly increased the cell perimeters, causing cell elongation, which is known to stimulate stress fiber activity and increase tractions [45,46]. To rule this out, we examined phase contrast images of cells at various time points after the CN03 treatment. The cell perimeters appeared to increase around 180-230 min after the treatment, whereas traction increased earlier, around 120 min after the treatment (Fig. S11). Data from multiple experiments showed that RMS traction and traction imbalance became statistically larger 120 and 180 min after the treatment, whereas perimeter became statistically larger than control 230 min after the treatment (Fig. 6e-g). Therefore, the CN03 increases the traction before the perimeter. The finding that CN03 increases traction before perimeter gives additional evidence supporting the notion that tractions, and the associated stress fibers that produce them, are the underlying factor affecting perimeter.

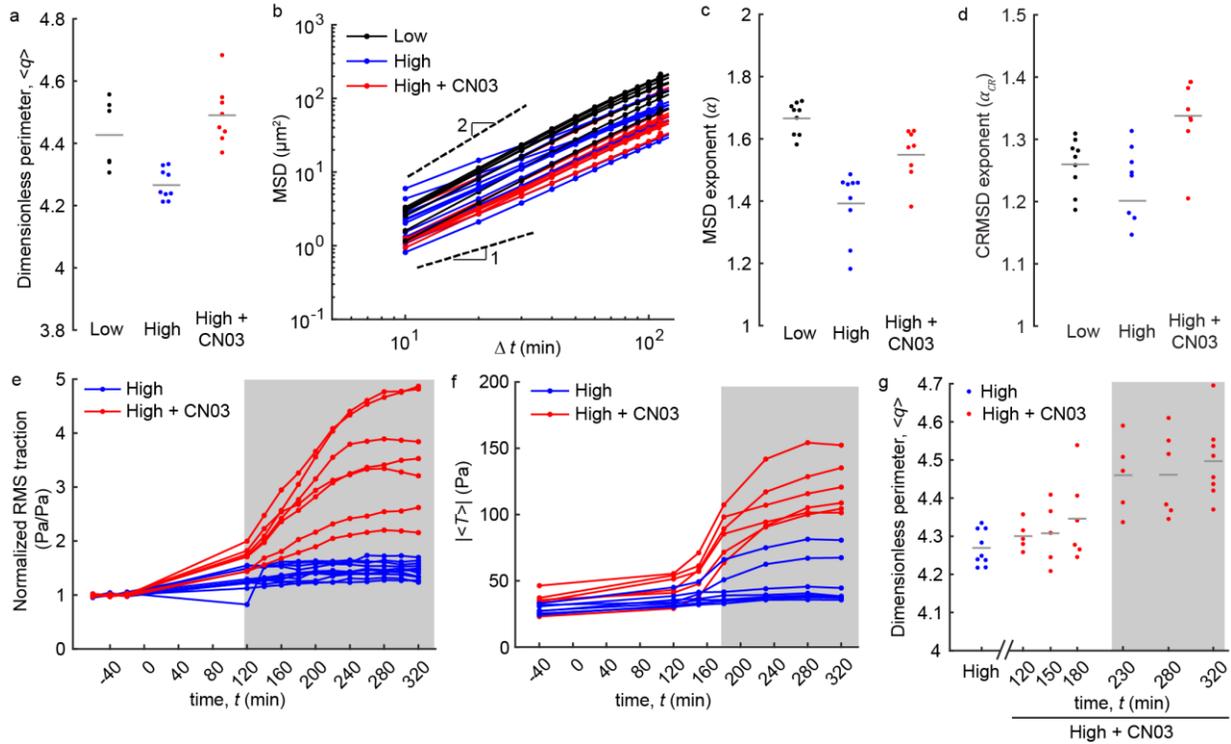

FIG. 6. Increasing cell-substrate tractions reverses effect of density on cell perimeter and cell rearrangements. (a) Average perimeter $<q>$ for cells in islands of low density (1200 cells/mm$^2$), high density (2600 cells/mm$^2$), or high density with the Rho activator CN03 (2 μg/mL). The low density and CN03 groups are statistically different from high density ($p < 0.005$). (b) Mean square displacement (MSD) for low density, high density, and CN03-treated high density islands. (c) Exponent $α$ of MSD for low density and CN03-treated high density islands are different from untreated high density islands ($p < 0.005$). (d) Exponent $α$ of CR-MSD for low density and CN03-treated high density islands is different from untreated high density islands ($p < 0.01$). (e) Average RMS traction for high density and CN03-treated high density islands measured over time before ($t < 0$) and after ($t > 0$) treatment with vehicle control or CN03. Shading shows time points for which tractions produced by CN03-treated cells are statistically different from high density control ($p < 0.001$). (f) Average traction imbalance for high density and CN03-treated high density islands measured over time before ($t < 0$) and after ($t > 0$) treatment with vehicle control or CN03 for a circular region of radius $R = 12$ μm. Shading shows timepoints for which CN03-treated cells are statistically different from high density control ($p < 0.001$). (g) Average cell perimeter becomes statistically larger than control at 230 minutes after the CN03 treatment ($p < 0.005$).

The Rho activator also significantly increased cell rearrangements in the high-density cell islands, as quantified by the exponents of MSD and CRMSD (Figs. 6c-d, S12). The difference was most notable in the exponent of CRMSD $α_{CR}$, which increased to a value above that of the low density islands (Fig. 6d). This observation can be explained by the fact that average cell displacement is lower at high density compared to low (Fig. 6b), which tends to increase the exponent of CRMSD for the high density islands compared to low. Together, the data demonstrate that by activating stress fibers to increase cell-substrate tractions, the effect of cell density on perimeter and rearrangements can be reversed.

## III. DISCUSSION

Though the relationship between force and migration in a cell monolayer is still poorly understood, a useful framework is the analogy between cell monolayers and glassy materials, specifically the transition between fluid-like and solid-like states [47]. Perhaps the strongest evidence for this analogy is the observation that cells appear to transition toward solid-like with increasing density [4,5,7,11–14]. Here, we hypothesized that changing density affected the motion by altering either the line tension at the cell-cell interface or the traction produced by the stress fibers. We observed increasing density to reduce tractions, average cell perimeter $<q>$, and cell rearrangements. Similarly, tractions, perimeters, and rearrangements were reduced by inhibitors of actomyosin contraction. In contrast to current understanding, the changes in perimeter and rearrangements appeared unrelated to the line tension associated with cortical tension and cell-cell adhesions. This led to the hypothesis that traction is the primary driver of cell rearrangements. To test this hypothesis, we activated actomyosin contraction in cell islands of high density, which caused cell perimeters and rearrangements to match those in monolayers at low density, thereby reversing the apparent effect of density on cell rearrangements. We conclude that perimeter and rearrangements in an adherent epithelial monolayer are controlled primarily by tractions produced by the stress fibers.

Our findings provide insights to interpret recent theories and experiments which showed that cell perimeter controls a transition between solid-like and fluid-like states in a cell monolayer [6,15,18–26]. These observations on the importance of perimeter have focused attention on the force-supporting components at the periphery of each cell, namely the cortex and the cell-cell adhesions, which together generate a line tension that acts like a surface tension [6,15,18–27]. Here we studied the line tension at each cell periphery by recoil velocity after laser ablation, imaging phosphorylated myosin in the cortex, and imaging E-cadherin at the cell-cell adhesions. According to current understanding [6,15,18–27], both increased recoil velocity and increased phosphorylated myosin would imply greater cortical tension, which would reduce the perimeter; increased E-cadherin would imply greater adhesion, which would increase the perimeter. Phosphorylated myosin and E-cadherin were unchanged by increasing cell density (Fig. 2), even as the perimeter was reduced (Fig. 1g). In separate experiments, blebbistatin and cytochalasin D each reduced the cortical tension (Fig. 3), which would be expected to increase the perimeter. By contrast, perimeter was reduced by these treatments (Fig. 3i-j). Together, these observations suggest that either current understanding on the relationship between line tension and perimeter is incorrect or some factor other than line tension at the cell periphery affects the cell perimeters. Our study suggests that cell perimeters are affected by another factor, traction.

The data show a strong connection between cell-to-substrate traction, aligned stress fibers, and cell perimeter. Though it is clear that stress fibers and traction are closely related, the underlying cause of the increased perimeter is unclear. It may be that increased cell perimeter results from the stress fibers, which may alter the internal forces within each cell, thereby polarizing that cell along a specific direction. Alternatively, the increased perimeter may result from the fact that imbalanced traction alters the balance of forces at each cell-cell interface, as studied by a theoretical model [21]. In this case, the relationship between perimeter and traction would be an emergent phenomenon, resulting from the balance of forces between cells rather than within each cell. This distinction is perhaps not so important, however, because either case points to the same result—greater tractions and greater forces in the stress fibers produce greater cell rearrangements within the cell layer.

The notion that stress fibers and traction are the key drivers of cell shapes and rearrangements in a monolayer provides insight into a previous observation in collective cell migration on the relationship between contraction and motion [6,28]. Greater contraction would presumably cause greater cortical tension, which would reduce both perimeter and rearrangements. By contrast, experiments observed greater contraction to be associated with larger perimeter and more cell rearrangements [6,28]. This observation was hypothesized to result from a change in the adhesion molecules between cells [6]. Our data suggests a different resolution—that the greater perimeter and cell rearrangements occurred not because of the forces at the cell-cell interfaces but rather because of the stress fibers and cell-substrate tractions. Indeed, greater cell rearrangement was observed previously to coincide with greater traction [6,28], which is consistent with a theoretical model that considered traction in addition to cortical tension and adhesion [21]. Our findings also provide insight into a recent study that proposed density to affect rearrangements by modulating cell-cell adhesions, which were suggested to create an effective friction that increased with density [7]. Though it is possible that friction could affect the migration, our finding that cell perimeters are governed primarily by stress fibers and traction rather than cortical tension or adhesion suggests that tractions produced by stress fibers are the primary driver of cell rearrangements.

A logical extension of the observation that cell-substrate tractions control the perimeter and rearrangement more so than cortical tension or cell-cell adhesion is that the energy associated with traction is greater than that associated with cortical tension or adhesion. Cells with large traction and large perimeter have well-established, highly-ordered stress fibers, which are present across the entire area of each cell (e.g., Fig. 5). By contrast, contraction in the cortex is present only at the periphery of each cell. As the area associated with the stress fibers is much larger than that associated with cortical tension, the magnitude of force produced in the stress fibers is likely to be far greater than in the cortex, causing the stress fibers, and therefore the tractions that they apply to the substrate, to be the dominant factor affecting cell perimeter and rearrangement. We expect this reasoning to apply to any epithelial cell type that applies large tractions to the substrate. Consistent with this, monolayers of human bronchial epithelial and MCF-10A cells appear to show relationships between tractions, perimeter, and collective migration [6,28]. Additionally, stress fiber activity is related to elongation of ovarian follicle cells in Drosophila development [48]. If there are no stress fibers, the cortical tension and cell-cell adhesions could control the cell perimeters and rearrangements. This may be the case in those aspects of development where there is little traction between cell and substrate [18,49–52]. Cell spheroids also have no cell-substrate tractions, and, consistent with our reasoning, cell perimeters are controlled by adhesions and cortical tension in cell spheroids [53].

Recent studies have proposed a jamming phase diagram with different axes representing the factors controlling whether or not cells rearrange positions with their neighbors [14,21,47,54]. Density has been proposed as one axis [14,47,54], but our data show that increasing the tractions can reverse the apparent effect of density on cell perimeter and rearrangements. This can be explained by a theoretical model that proposed an additional axis of the phase diagram proportional to traction [21]. Other theoretical models have suggested that the axes of cell density and contraction are coupled but nevertheless separate [14,54]. Although we do not rule out the possibility that cell density is a distinct axis on the phase diagram, the experiments here demonstrate that effects of cell density on cell shape and rearrangements can be explained largely by cell traction produced by stress fibers. Another proposed axis of the phase diagram is related to the line tension $\Gamma$, resulting from the balance of adhesion and cortical tension [6,21]. Though we similarly do not rule out the existence of this axis, our results demonstrate that the cell perimeters and rearrangements are far more sensitive to traction. We expect that in systems where tractions are small, such as an embryo, the situation may be reversed, with adhesion and cortical tension playing a dominant role. It is likely that still other axes of the phase diagram await to be discovered. The importance of stress fibers and traction observed here provides a starting point for future experiments and answer these questions.

## IV. MATERIALS AND METHODS

### A. Cell culture

Madin-Darby canine kidney type II cells, expressing either GFP with a nuclear localization signal or Lifeact-GFP were supplied by the laboratory of Prof. David Weitz, Harvard University. The cells were maintained in low-glucose Dulbecco's modified Eagle's medium (12320-032; Life Technologies, Carlsbad, CA) with 10% fetal bovine serum (Corning, NY) and 1% G418 (Corning) in an incubator at 37ºC and 5% $CO_2$. Cells with nuclear GFP were used for traction, migration, and immunofluorescence experiments, whereas Lifeact cells were used for laser ablation experiments. For experiments using blebbistatin, cytochalasin D, and SMIFH2, cell media was replaced with 2% fetal bovine serum 3-4 hours before the treatments. For experiments with CN03, cell media was replaced with 1% fetal bovine serum one day before the treatment. Time lapse experiments were performed for 4-6 hours to maintain a relatively constant density.

### B. Polyacrylamide substrates

Polyacrylamide gels with Young's modulus of 6 kPa and thickness of 150 μm were prepared with fluorescent particles located at the top. A gel solution of 5.5% weight/volume (w/v) acrylamide (Biorad Laboratories, Hercules, CA), 0.2% w/v bisacrylamide (Biorad) was prepared, and 20 μL was pipetted onto #1.5 glass bottom dishes (Cellvis, Mountain View, CA). A glass coverslip (18-mm diameter circle) was placed on top of each gel and removed after the gel solution was polymerized. Then, a second gel solution with composition of the first gel plus 0.036% w/v fluorescent particles (diameter 0.5 μm, carboxylate-modified; Life Technologies) was prepared, and 20 μL was pipetted on the polymerized first gel. Again, a coverslip was placed on top, and the dishes were centrifuged upside down to localize the fluorescent particles to the top of the second gel. The top surface of the second gel was functionalized with type I rat tail collagen (BD Biosciences, Franklin Lakes, NJ; 0.01 mg/mL, 1-2 mL per 18-mm diameter gel) using the covalent cross-linker sulfo-SANPAH (Pierce Biotechnology, Waltham, MA).

**C. Micropatterning confined cellular islands**

Polydimethysiloxane (PDMS) (Sylgard 184, Dow Corning, Midland, MI) was poured onto plastic dishes to cure for 4 hours on a hot plate at 70ºC to make 400-600 μm thick sheets. These sheets were cut into 16 mm circular masks, and biopsy punches (1 mm diameter) were used to make holes in the PDMS masks. The masks were sterilized with 70% ethanol and incubated overnight at room temperature in 2% Pluronic F-127 (Sigma-Aldrich, St. Louis, MO) to prevent cell adhesion to the masks. Masks were placed on the polyacrylamide gels before functionalizing with sulfo-SANPAH and collagen, thereby constraining the collagen to circular patterns on the gels. 300 μL of cell solution of concentration 0.5 million cells/mL was pipetted onto the masks and incubated at 37ºC for 2 hours. The masks were then removed, and the patterned cell islands were placed in the incubator at 37ºC and 5% $CO_2$ until they attained the desired density for imaging.

**D. Widefield microscopy**

Images of the cell islands and the fluorescent particles were captured every 5 or 10 minutes using phase contrast and fluorescent modes of an Eclipse Ti microscope (Nikon, Melville, NY) with a 10× numerical aperture 0.5 objective and an Orca Flash 4.0 digital camera (Hamamatsu, Bridgewater, NJ) running Elements Ar software (Nikon). The imaging environment was maintained at 37°C and 5% $CO_2$ using a H301 stage top incubator with UNO controller (Okolab USA Inc, San Bruno, CA). After the time lapse experiments, cells were removed from the polyacrylamide substrates by incubating in 0.05% trypsin for 20 minutes, and images of the fluorescent particles were collected; these images provided a traction-free reference state for computing cell-substrate tractions.

### E. Traction force microscopy

The term "traction" in this manuscript refers to the vector field of in-plane force per area applied by the cells to the substrate. To quantify tractions, cell-induced displacements of the fluorescent particles were measured using Fast Iterative Digital Image Correlation (FIDIC) [55] using 32x32 pixel subsets centered on a grid with a spacing of 8 pixels (5 µm). Tractions were subsequently computed using unconstrained Fourier transform traction microscopy [38] accounting for the finite substrate thickness [39,40]. The computed traction has components in the $x$ and $y$ directions $T_x$ and $T_y$, respectively. The root-mean-square (RMS) of traction was computed according to the expression $[<T_x^2 + T_y^2>]^{1/2}$, where the angle brackets $<>$ indicate a mean over all positions. The average traction imbalance was computed for each grid point according to the expression $|\sum_i(\vec{T}_i A)/\sum_i A|$, where $\vec{T}_i$ is the traction vector, $A$ is the (constant) area over which each traction vector acts, and the sum is a vector sum over all locations $i$ within a circle of radius $R$ from the grid point of interest. The computation was repeated for $R$ ranging from 9 to 90 µm. This was repeated for every grid point at least 60 µm from the edge of each cell island and averaged over the cell island. The symbol $|<T>|$ is used to represent the average traction imbalance.

### F. Cell velocities and trajectories

Cell velocities were measured using FIDIC [55] from phase contrast images of cell islands. Consecutive images were correlated, and the resulting displacements were divided by time to compute velocity. Subsets of 48x48 pixels were used with a spacing of 12 pixels (8 µm). The cell island boundaries were detected using Matlab 2015a code written based on ref. [56].

The cell displacements computed with FIDIC were used to calculate cell trajectories as described in refs. [6,28]. The calculation began with an initial evenly spaced grid, with the distance between grid points matching the average cell size. Displacements from FIDIC were interpolated to each grid point, and the interpolated values were added to the coordinates of each grid point to give an estimate of each cell's position at the next time point. This process was repeated for each time point of the experiment, thereby creating a list of points for each cell corresponding to that cell's trajectory.

**G. Mean square displacements**

As a metric for cell rearrangement, the mean square displacement (MSD) was computed according to the equation

$$MSD\,(\Delta t) = <|u_i(t,\Delta t)|^2> = <|r_i(t+\Delta t) - r_i(t)|^2> \qquad (Eq.\ 1a)$$

where $r_i(t)$ is the position of the $i^{th}$ cell at time $t$, $u_i(t,\Delta t)$ is the displacement of the $i^{th}$ cell for time interval $\Delta t$ and the brackets <> denote an average over all cells and all possible starting time points $t$. In some figures of the supplementary information, the mean square displacement is shown for individual cells, in which case the average is taken over all $t$; for these figures, the notation $MSD_i$ is used to indicate that the curves represent individual cells. As tractions were nearly constant for a time span of 120 min, the MSD was computed over a time span of 120 min.

**H. Cage relative mean square displacements**

As a second metric for rearrangements, cage-relative mean square displacement (CRMSD) was computed as explained in ref. [29]. The CRMSD quantifies each cell's motion relative to the local motion of its $N$ nearest neighbors according to the equation

$$CRMSD\,(\Delta t) = <|u_i(t,\Delta t) - u_i^{cage}(t,\Delta t)|^2> \qquad (Eq.\ 1b)$$

where $u_i^{cage}(t,\Delta t)$ is the average displacement of the nearest $N$ neighbors to cell $i$ and $N$ was equal to 4, 10, or 20, as described in the text. The CRMSD was computed over the same time span as the MSD.

**I. Chemical treatments**

Chemical treatments were blebbistatin (Sigma-Aldrich), cytochalasin D (Sigma-Aldrich), and CN03 (Cytoskeleton, Inc, Denver, CO). Stock solutions of blebbistatin, cytochalasin D, and CN03 were prepared at 20 mM, 2 mM, and 0.1 g/L respectively, all dissolved in DMSO except CN03 in water. The stock solutions were diluted in PBS to obtain desired concentrations for the experiments.

**J. Immunofluorescence and confocal microscopy**

For immunofluorescent labeling, MDCK cells were seeded to form monolayers/islands of required confluence and treated with the chemical of interest. After 3 hours, the cells were rinsed twice with PBS and fixed with 4% paraformaldehyde in PBS for 20 minutes. The cells were washed with Tris-buffered saline twice for 5 minutes each. For actin staining, cells were incubated in 0.1% Triton X-100 for 5 minutes at room temperature, and then placed in 3-5 units/mL solution of Phallodin Dylight 594 (Life Technologies catalog no. 21836). For E-cadherin and myosin staining, cells were treated with E-Cadherin rabbit antibody (1:200 ratio; Cell Signaling, Danvers, MA, catalog no. 24E10) or phospho-Myosin Light Chain 2 (Ser 19) rabbit antibody (1:100 ratio; Cell Signaling catalog no. 3671). After washing the cells, Alexa 647 anti-rabbit secondary antibody (1:400 ratio; Life Technologies, catalog no. A-21245) was used for fluorescent staining.

Fluorescently labeled cells were imaged with a Nikon A1R+ confocal microscope with a 40× NA 1.15 water-immersion objective with a step size of 0.5 μm using Elements Ar software (Nikon). For imaging cortical actin, E-cadherin, and phosphorylated myosin light chain (pMLC), image stacks were captured near the apex of the cells, and a maximum intensity projection having a thickness of 3 μm was computed. For imaging stress fibers, image stacks were captured at the base of the cells, and a maximum intensity projection having thickness of 1.5 μm was computed. Representative images shown in the figures were cropped and pseudocolored using ImageJ.

**K. Measurement of dimensionless cell perimeter**

A maximum intensity projection of the confocal images of F-actin near the apex of the cell layer gave images of the cell boundaries. The cells were then segmented using Seedwater Segmenter [57], and coordinates of each cell boundary were identified using the bwboundaries function in Matlab 2015a. Vertices, defined as locations where 3 or more cells met, were identified from the coordinates of the cell boundaries. Each cell perimeter was computed by summing the distances between the vertices. Each cell area was computed from the identified vertices using the polyarea function in Matlab 2015a. The dimensionless cell perimeter $q$ was then computed according to $q = PA^{-1/2}$. Finally, the mean of $q$, $<q>$, was computed over 100-150 cells.

**L. E-cadherin and pMLC analysis**

Fluorescent intensities of E-cadherin and pMLC were quantified in the cell cortex, located at the cell peripheries. To account for background signal, the median of each image was subtracted from that image. Peripheral locations were defined by dilating the segmented images used to measure cell perimeter by 1 µm; pixels within the dilated images were defined as being at the cell peripheries and used to quantify intensities of E-cadherin and pMLC. For comparisons between control and treatments, the intensities were normalized by the average of the control.

**M. Order parameter**

Stress fibers were quantified within the bulk of each cell. To determine positions for analysis, the same segmentation as used for E-cadherin and pMLC was used, except stress fibers were defined to be outside the dilated cell boundaries. We quantified actin stress fiber alignment in each cell using an order parameter $S$. For this, we calculated the angle $\theta$ at each pixel inside a cell of interest using the OrientationJ plugin [58] in ImageJ and computed the average of the angles, $\theta_m$. The order parameter $S$ for the cell is defined as

$$S = 2(<cos^2(\theta - \theta_m)> - 1/2) \quad \text{(Eq. 2)}$$

where the brackets <> indicate an average over all positions within the cell. The order parameter $S$ attains a value of 1 for fully aligned stress fibers and 0 for random alignment.

**N. Laser ablation**

Laser ablation was performed on a CSU-X spinning disk confocal microscope (Yokogawa) mounted on an Eclipse Ti base with a 60× NA 1.4 oil immersion objective (Nikon) and imaged with a Zyla sCMOS camera (Andor), all run by the IQ3 acquisition software (Andor). MDCK Lifeact cells were seeded to confluence on glass substrates as in ref. [31], and cell-cell edges were visualized with the 488 nm laser line. A Micropoint laser at 405 nm was used to ablate cell-cell edges using a single 0.06 second pulse at 65% laser power. An image was captured before ablation to determine the initial distance $d(0)$, and images were captured every 5 s from 1 to 41 s after the ablation. The distance $d(t)$ between vertices of each ablated cell edge was measured manually for each time point. The initial recoil velocity was defined as the slope of the plot of $d(t) - d(0)$ vs. time from 0 to 11 s, which was computed using a least-mean-square fitting that was constrained to go through the origin.

**O. Statistical analysis**

Unless otherwise noted, each colored dot or line in the plots is an average taken over a different cell island or monolayer and represents an independent biological sample. Tractions, MSD, and CRMSD were measured in cell islands, whereas cell perimeter, fluorescent intensities, and order parameter were computed in larger cell monolayers. Statistical comparisons between groups were performed using a two-sided Student's t-test or, for multiple groups, one-way ANOVA with Tukey's correction for multiple comparisons. Analyses were performed in Matlab 2015a.

**ACKNOWLEDGMENTS**

We thank B. Burkel and M. Graham for comments on the manuscript. Confocal microscopy was performed at the Biochemistry Optical Core and the Materials Science Center of the University of Wisconsin-Madison. This work was supported by National Science Foundation grant number CMMI-1660703.

**APPENDIX: SUPPLEMENTAL FIGURES**

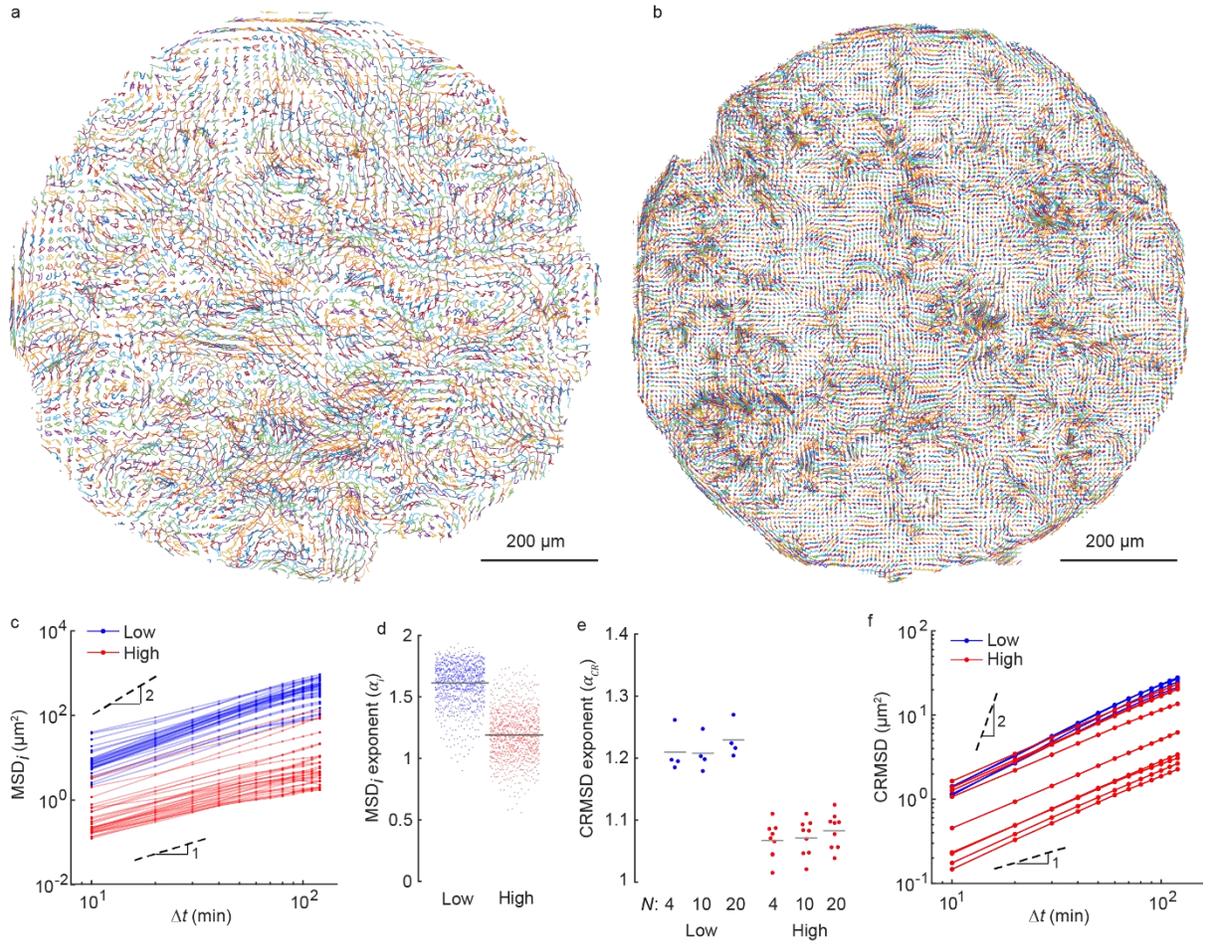

FIG. S1. Effect of density on collective motion. Cell trajectories computed over a time span of 8 hours in low (1200 cells/mm$^2$) (a) and high (4200 cells/mm$^2$) (b) density islands. These are the same cell islands as in Fig. 1a, c. (c) Mean square displacement computed from trajectories of selected cells (MSD$_i$) from single islands of low and high density. (d) Exponent $\alpha_i$ of the MSD$_i$ curves in panel c ($p < 0.001$). Each dot in panel d represents an individual cell. (e) Exponent $\alpha_{CR}$ of the CRMSD curves for different cell pack sizes, $N$, for cell islands of low and high densities. (f) CRMSD of low and high density islands for $N = 20$.

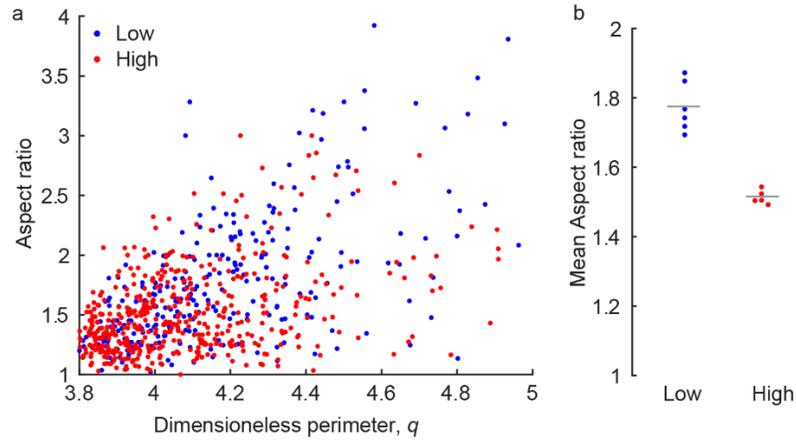

FIG. S2. Correlation between aspect ratio and dimensionless perimeter. (a) Scatter plot between aspect ratio and dimensionless perimeter for cells at low (1200 cells/mm$^2$) and high density (4200 cells/mm$^2$). Each dot corresponds to a different cell. Aspect ratio was computed from the cell vertices using the regionprops command of Matlab 2015a. (b) Mean aspect ratio is smaller at high density compared to low density ($p < 0.001$). Each dot corresponds to an average of 100-150 cells in a different field of view.

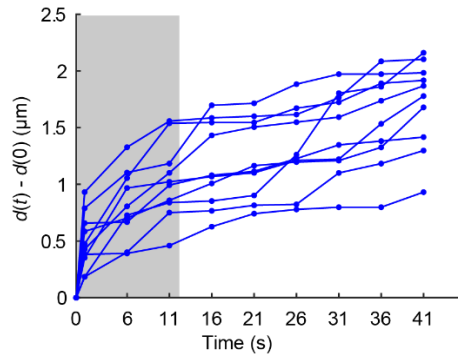

FIG. S3. Recoil in vertices that connect a cell edge after laser ablation. Change in distance, $d(t) - d(0)$, between vertices after laser ablation of cells in a monolayer of low density (1200 cells/mm$^2$). Initial recoil velocity is computed by fitting a line to data points in gray region, corresponding to $t = 0$–11 s. Each line corresponds to an ablated cell edge.

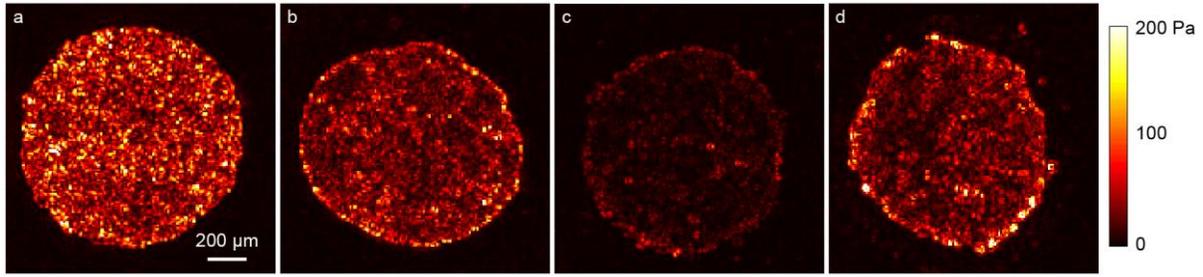

FIG. S4. (a-b) Colormaps of traction magnitude in low density (1200 cells/mm$^2$) (a), high density (4200 cells/mm$^2$) (b), blebbistatin-treated (1200 cells/mm$^2$) (c), and cytochalasin D-treated (1200 cells/mm$^2$) (d) islands.

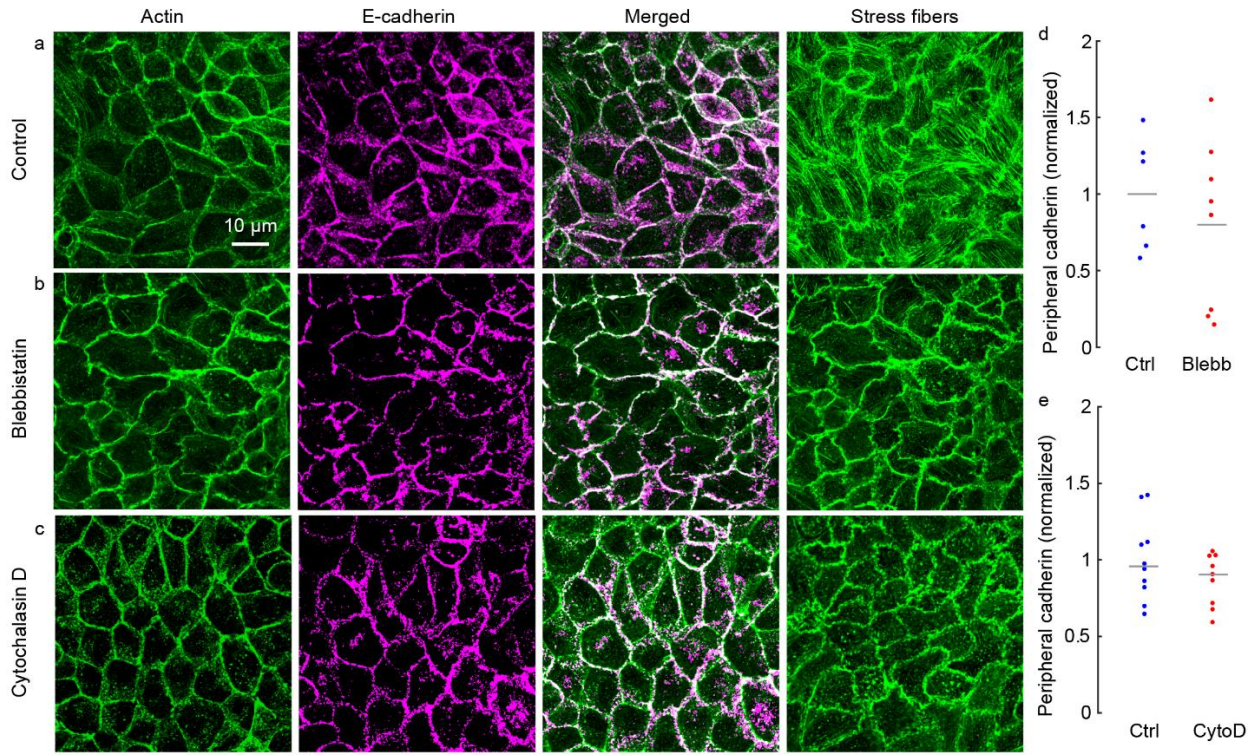

FIG. S5. Effect of blebbistatin and cytochalasin D on E-cadherin and stress fibers. (a-c) Confocal images of cortical actin, E-cadherin, and actin stress fibers for control cell monolayers (a) and cell monolayers treated with 20 μM blebbistatin (b) or 0.05 μM cytochalasin D (c). (d, e) Peripheral E-cadherin intensities are unaffected by 20 μM blebbistatin (d, $p = 0.45$) or 0.05 μM cytochalasin D (e, $p = 0.23$). The cell density in all panels is 1200 cells/mm$^2$.

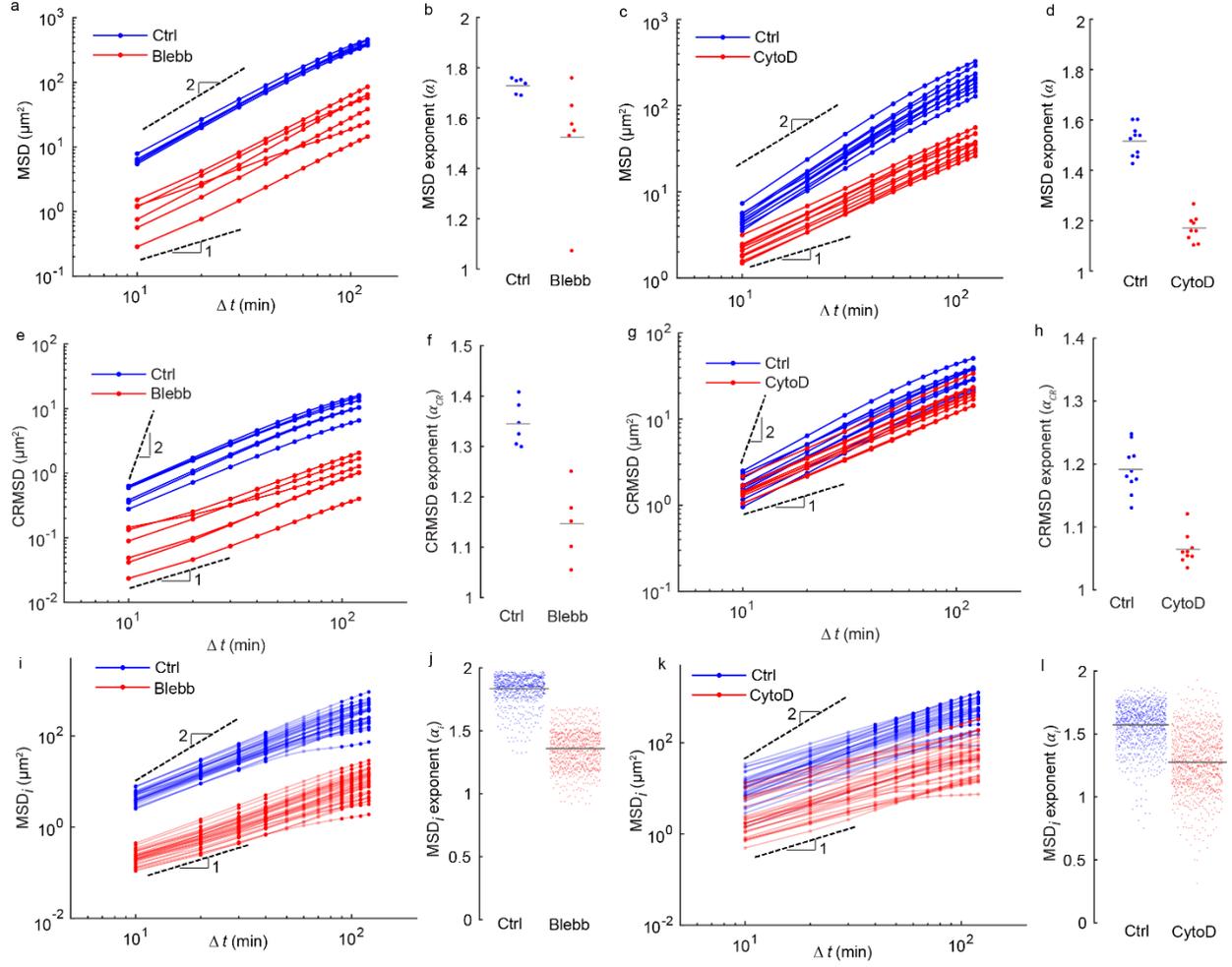

FIG. S6. Cell rearrangements decrease with inhibition of actomyosin contractility. (a, c) MSD of control cell islands and islands treated with 20 μM blebbistatin (a) or 0.05 μM cytochalasin D (c). (b, d) Exponent a of MSD decreases after treating with 20 μM blebbistatin (b, $p < 0.05$) or 0.05 μM cytochalasin D (d, $p < 0.001$). (e, g) CRMSD of control cell islands and islands treated with 20 μM blebbistatin (e) or 0.05 μM cytochalasin D (g). (f, h) Exponent $\alpha_{CR}$ of CRMSD decreases after treating with 20 μM blebbistatin (f, $p < 0.001$) or 0.05 μM cytochalasin D (h, $p < 0.001$). (i, k) $MSD_i$ for selected cell trajectories in control islands and islands treated with 20 μM blebbistatin (i) or 0.05 μM cytochalasin D (k). (j, l) Exponent $\alpha_i$ of $MSD_i$ of cell trajectories in control islands and islands treated with 20 μM blebbistatin (j, $p < 0.001$) or control islands and islands treated with 0.05 μM cytochalasin D (l, $p < 0.001$). Dots in panels j and l represent individual cells. The cell density in all panels is 1200 cells/mm$^2$.

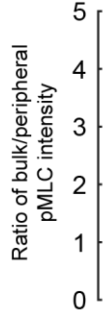

FIG. S7. Ratio of total pMLC intensity in bulk and peripheral region. Peripheral pMLC was quantified as described in the methods. Bulk pMLC was quantified in the same region as the stress fibers, namely at the base of the cell layer and away from the cell peripheries. The cell density is 1200 cells/mm$^2$.

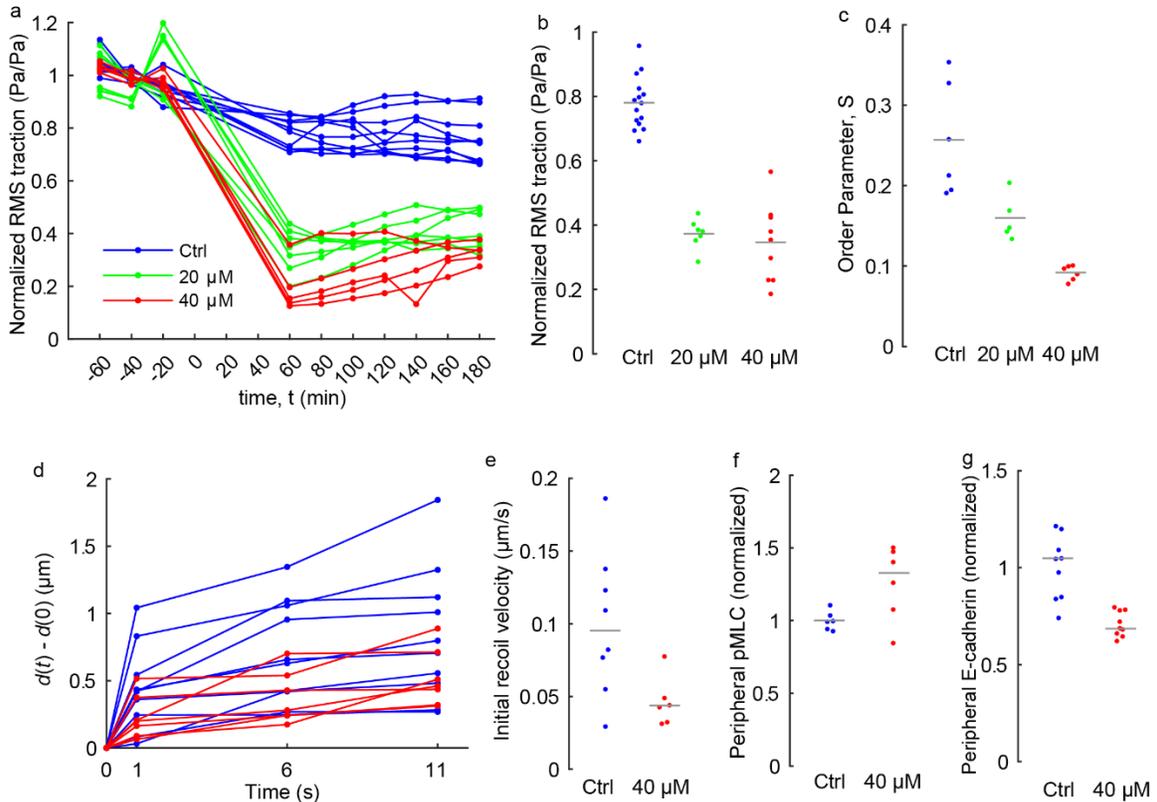

FIG. S8. Stress fiber inhibition. (a) Average RMS traction for each island is measured over time before ($t < 0$) and after ($t > 0$) treating with the formin inhibitor SMIFH2. For each island, the RMS traction is normalized by its average value before the treatment. (b) Normalized RMS traction for $t = 60$–180 min. Relative to control, both concentrations reduce the RMS traction ($p < 0.001$). (c) Order parameter $S$ decreases after stress fiber inhibition ($p < 0.001$). The cell density is 1200 cells/mm$^2$. (d) Change in distance between the vertices of an ablated edge, $d(t) - d(0)$, for control and cells treated with 40 μM SMIFH2. (e) Initial recoil velocity after laser ablation is decreased by SMIFH2 ($p < 0.01$). (f) Peripheral pMLC fluorescent intensity is increased by SMIFH2 ($p < 0.05$). (g) Peripheral E-cadherin intensity is decreased by SMIFH2 ($p < 0.01$).

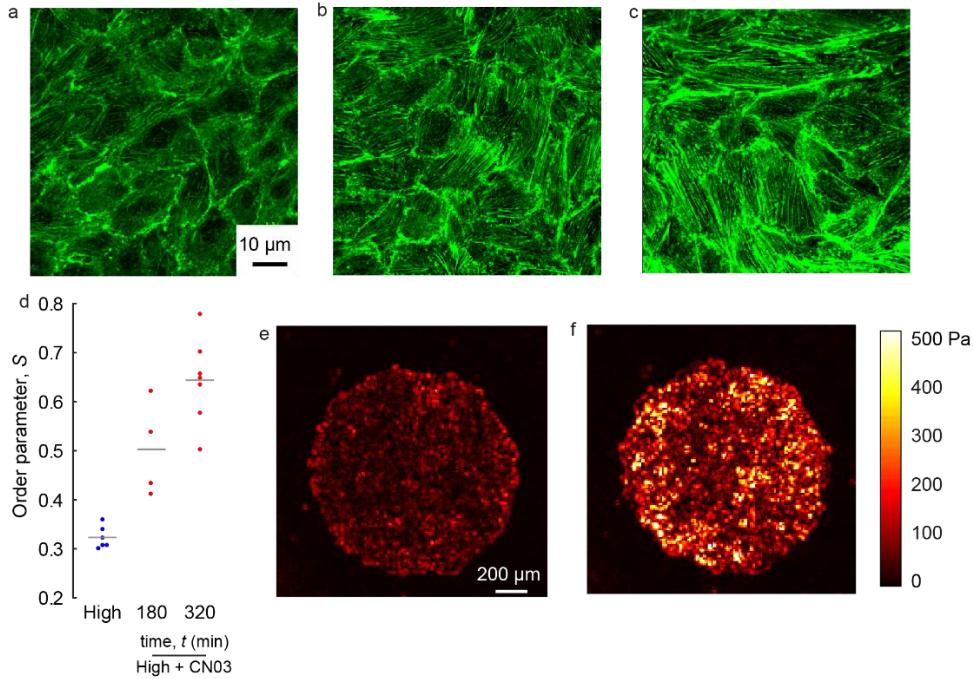

FIG. S9. Rho activator CN03 increases stress fibers and tractions. (a–c) Confocal images of actin stress fibers for control cells at high density (a) and cells at high density at $t = 180$ min (b) or 320 min (c) after treating with CN03 (2 μg/mL). (d) Order parameter for high density is smaller than for high density treated with CN03 ($p < 0.001$ for both groups compared to high density control). (e, f) Colormaps of traction magnitude of high density (e) and high density treated with CN03 at $t = 320$ min (f). Cell density for all data in this figure is 2600 cells/mm$^2$.

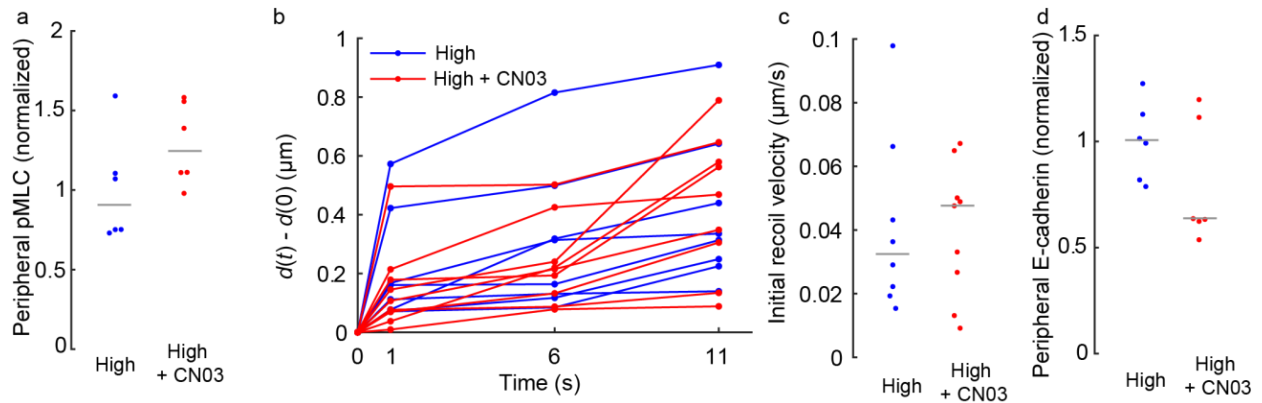

FIG. S10. Effect of CN03 (2 μg/mL) on line tension. (a) Fluorescent intensity of peripheral pMLC ($p = 0.12$). (b, c) Change in distance $d(t) - d(0)$ after laser ablation (b) and initial recoil velocity ($p = 0.86$, c). (d) Fluorescent intensity of peripheral E-cadherin ($p = 0.15$). Cell density for all data in this figure is 2600 cells/mm$^2$.

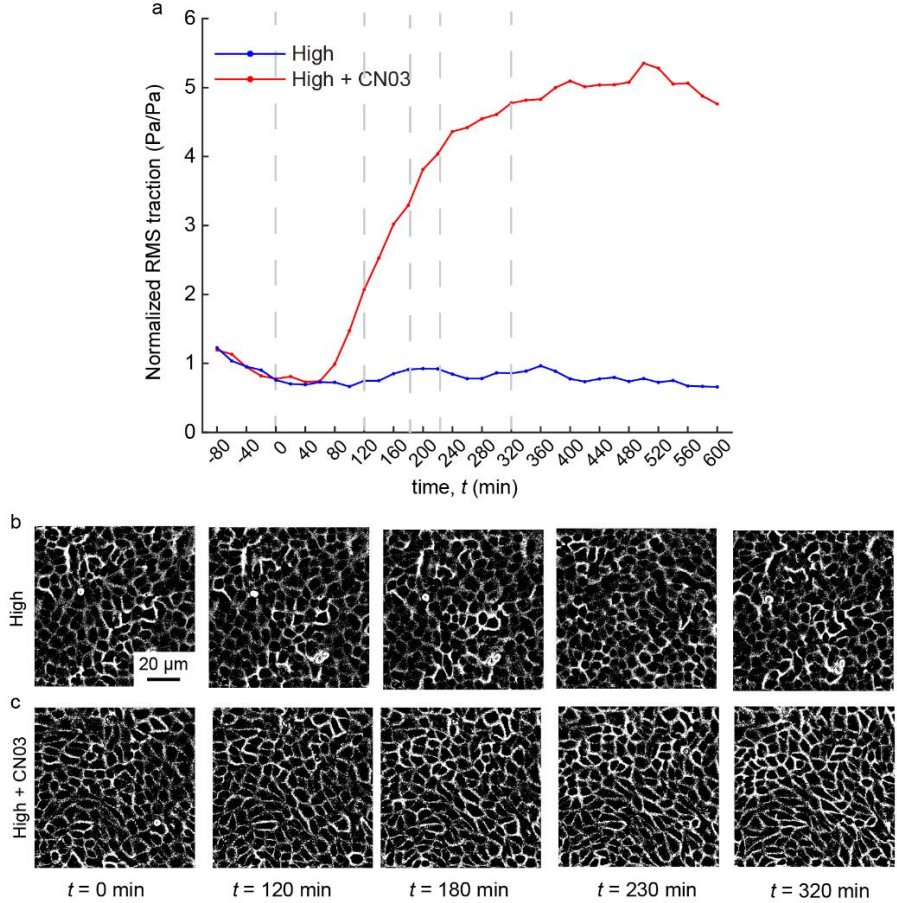

FIG. S11. Changes in traction precede changes in dimensionless cell perimeter. (a) Average RMS traction for high density island (2600 cells/mm$^2$) and high density islands treated with CN03 (2 µg/mL) measured over time before ($t < 0$) and after ($t > 0$) treatment with vehicle control or CN03. (b, c) Phase contrast images of control cells (b) and CN03-treated cells (c) at different times during the treatment.

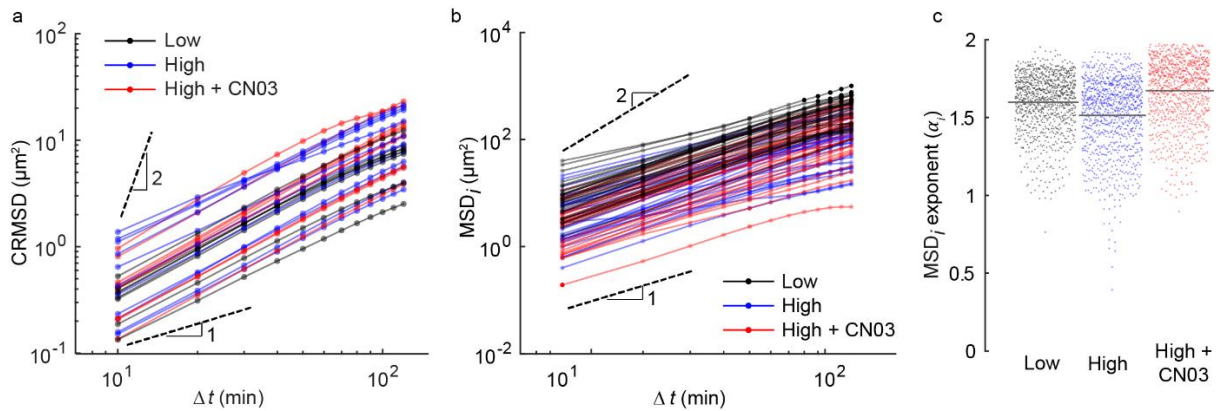

FIG. S12. Rho activator CN03 increases cell rearrangements. (a) CRMSD for islands of low density, high density, and high density treated with CN03 (2 µg/mL). (b) MSD$_i$ for selected cell trajectories in islands of low density (1200 cells/mm$^2$), high density (2600 cells/mm$^2$), and high density treated with CN03 islands (2 µg/mL). (c) Exponent $\alpha_i$ of MSD$_i$ of cell trajectories in islands of low density, high density, and high density treated with CN03 (2 µg/mL). Each dot represents an individual cell.

**REFERENCES**

1. Friedl P, Gilmour D. 2009 Collective cell migration in morphogenesis, regeneration and cancer. *Nature Reviews Molecular Cell Biology* **10**, 445–457. (doi:10.1038/nrm2720)

2. Oswald L, Grosser S, Smith DM, Käs JA. 2017 Jamming transitions in cancer. *Journal of Physics D: Applied Physics* **50**, 483001. (doi:10.1088/1361-6463/aa8e83)

3. Mongera A *et al.* 2018 A fluid-to-solid jamming transition underlies vertebrate body axis elongation. *Nature* **561**, 401–405. (doi:10.1038/s41586-018-0479-2)

4. Angelini TE, Hannezo E, Trepat X, Marquez M, Fredberg JJ, Weitz DA. 2011 Glass-like dynamics of collective cell migration. *Proceedings of the National Academy of Sciences* **108**, 4714–4719. (doi:10.1073/pnas.1010059108)

5. Nnetu KD, Knorr M, Pawlizak S, Fuhs T, Käs JA. 2013 Slow and anomalous dynamics of an MCF-10A epithelial cell monolayer. *Soft Matter* **9**, 9335. (doi:10.1039/c3sm50806d)

6. Park J-A *et al.* 2015 Unjamming and cell shape in the asthmatic airway epithelium. *Nature Materials* **14**, 1040–1048. (doi:10.1038/nmat4357)

7. Garcia S, Hannezo E, Elgeti J, Joanny J-F, Silberzan P, Gov NS. 2015 Physics of active jamming during collective cellular motion in a monolayer. *Proceedings of the National Academy of Sciences* **112**, 15314–15319. (doi:10.1073/pnas.1510973112)

8. Atia L *et al.* 2018 Geometric constraints during epithelial jamming. *Nature Physics* **14**, 613–620. (doi:10.1038/s41567-018-0089-9)

9. Liu AJ, Nagel SR. 1998 Jamming is not just cool any more: Nonlinear dynamics. *Nature* **396**, 21–22. (doi:10.1038/23819)

10. Trappe V, Prasad V, Cipelletti L, Segre PN, Weitz DA. 2001 Jamming phase diagram for attractive particles. *Nature* **411**, 772–775. (doi:10.1038/35081021)

11. Tambe DT *et al.* 2011 Collective cell guidance by cooperative intercellular forces. *Nature Materials* **10**, 469–475. (doi:10.1038/nmat3025)

12. Puliafito A, Hufnagel L, Neveu P, Streichan S, Sigal A, Fygenson DK, Shraiman BI. 2012 Collective and single cell behavior in epithelial contact inhibition. *Proceedings of the National Academy of Sciences* **109**, 739–744. (doi:10.1073/pnas.1007809109)

13. Doxzen K, Vedula SRK, Leong MC, Hirata H, Gov NS, Kabla AJ, Ladoux B, Lim CT. 2013 Guidance of collective cell migration by substrate geometry. *Integrative Biology* **5**, 1026. (doi:10.1039/c3ib40054a)

14. Loza AJ, Koride S, Schimizzi GV, Li B, Sun SX, Longmore GD. 2016 Cell density and actomyosin contractility control the organization of migrating collectives within an epithelium. *Molecular Biology of the Cell* **27**, 3459–3470. (doi:10.1091/mbc.E16-05-0329)

15. Bi D, Lopez JH, Schwarz JM, Manning ML. 2015 A density-independent rigidity transition in biological tissues. *Nature Physics* **11**, 1074–1079. (doi:10.1038/nphys3471)

16. Vincent R, Bazellières E, Pérez-González C, Uroz M, Serra-Picamal X, Trepat X. 2015 Active Tensile Modulus of an Epithelial Monolayer. *Physical Review Letters* **115**. (doi:10.1103/PhysRevLett.115.248103)

17. Notbohm J *et al.* 2016 Cellular Contraction and Polarization Drive Collective Cellular Motion. *Biophysical Journal* **110**, 2729–2738. (doi:10.1016/j.bpj.2016.05.019)

18. Farhadifar R, Röper J-C, Aigouy B, Eaton S, Jülicher F. 2007 The Influence of Cell Mechanics, Cell-Cell Interactions, and Proliferation on Epithelial Packing. *Current Biology* **17**, 2095–2104. (doi:10.1016/j.cub.2007.11.049)

19. Staple DB, Farhadifar R, Röper J-C, Aigouy B, Eaton S, Jülicher F. 2010 Mechanics and remodelling of cell packings in epithelia. *The European Physical Journal E* **33**, 117–127. (doi:10.1140/epje/i2010-10677-0)

20. R. Noppe A, Roberts AP, Neufeld Z, Yap AS, Gomez GA. 2015 Modelling wound closure in an epithelial cell sheet using the cellular Potts model. *Integrative Biology* **7**, 1253–1264. (doi:10.1039/c5ib00053j)

21. Bi D, Yang X, Marchetti MC, Manning ML. 2016 Motility-Driven Glass and Jamming Transitions in Biological Tissues. *Physical Review X* **6**. (doi:10.1103/PhysRevX.6.021011)

22. Chiang M, Marenduzzo D. 2016 Glass transitions in the cellular Potts model. *EPL (Europhysics Letters)* **116**, 28009. (doi:10.1209/0295-5075/116/28009)

23. Barton DL, Henkes S, Weijer CJ, Sknepnek R. 2017 Active Vertex Model for cell-resolution description of epithelial tissue mechanics. *PLOS Computational Biology* **13**, e1005569. (doi:10.1371/journal.pcbi.1005569)

24. Yang X, Bi D, Czajkowski M, Merkel M, Manning ML, Marchetti MC. 2017 Correlating cell shape and cellular stress in motile confluent tissues. *Proceedings of the National Academy of Sciences* **114**, 12663–12668. (doi:10.1073/pnas.1705921114)

25. Moshe M, Bowick MJ, Marchetti MC. 2018 Geometric Frustration and Solid-Solid Transitions in Model 2D Tissue. *Physical Review Letters* **120**. (doi:10.1103/PhysRevLett.120.268105)

26. Czajkowski M, Bi D, Manning ML, Marchetti MC. 2018 Hydrodynamics of shape-driven rigidity transitions in motile tissues. *Soft Matter* **14**, 5628–5642. (doi:10.1039/C8SM00446C)

27. Brodland GW. 2002 The Differential Interfacial Tension Hypothesis (DITH): A Comprehensive Theory for the Self-Rearrangement of Embryonic Cells and Tissues. *Journal of Biomechanical Engineering* **124**, 188. (doi:10.1115/1.1449491)

28. Malinverno C *et al.* 2017 Endocytic reawakening of motility in jammed epithelia. *Nature Materials* **16**, 587–596. (doi:10.1038/nmat4848)

29. Lin S-Z, Ye S, Xu G-K, Li B, Feng X-Q. 2018 Dynamic Migration Modes of Collective Cells. *Biophysical Journal* **115**, 1826–1835. (doi:10.1016/j.bpj.2018.09.010)

30. Hutson MS. 2003 Forces for Morphogenesis Investigated with Laser Microsurgery and Quantitative Modeling. *Science* **300**, 145–149. (doi:10.1126/science.1079552)

31. Verma S *et al.* 2012 A WAVE2–Arp2/3 actin nucleator apparatus supports junctional tension at the epithelial zonula adherens. *Molecular Biology of the Cell* **23**, 4601–4610. (doi:10.1091/mbc.e12-08-0574)

32. Curran S, Strandkvist C, Bathmann J, de Gennes M, Kabla A, Salbreux G, Baum B. 2017 Myosin II Controls Junction Fluctuations to Guide Epithelial Tissue Ordering. *Developmental Cell* **43**, 480-492.e6. (doi:10.1016/j.devcel.2017.09.018)

33. Choi W, Acharya BR, Peyret G, Fardin M-A, Mège R-M, Ladoux B, Yap AS, Fanning AS, Peifer M. 2016 Remodeling the zonula adherens in response to tension and the role of afadin in this response. *J Cell Biol* **213**, 243–260. (doi:10.1083/jcb.201506115)

34. Staddon MF, Cavanaugh KE, Munro EM, Gardel ML, Banerjee S. 2019 Mechanosensitive Junction Remodeling Promotes Robust Epithelial Morphogenesis. *Biophysical Journal* **117**, 1739–1750. (doi:10.1016/j.bpj.2019.09.027)

35. Miroshnikova YA *et al.* 2018 Adhesion forces and cortical tension couple cell proliferation and differentiation to drive epidermal stratification. *Nature Cell Biology* **20**, 69–80. (doi:10.1038/s41556-017-0005-z)

36. Krieg M, Arboleda-Estudillo Y, Puech P-H, Käfer J, Graner F, Müller DJ, Heisenberg C-P. 2008 Tensile forces govern germ-layer organization in zebrafish. *Nature Cell Biology* **10**, 429–436. (doi:10.1038/ncb1705)

37. Engl W, Arasi B, Yap LL, Thiery JP, Viasnoff V. 2014 Actin dynamics modulate mechanosensitive immobilization of E-cadherin at adherens junctions. *Nature Cell Biology* **16**, 584–591. (doi:10.1038/ncb2973)

38. Butler JP, Tolić-Nørrelykke IM, Fabry B, Fredberg JJ. 2002 Traction fields, moments, and strain energy that cells exert on their surroundings. *American Journal of Physiology-Cell Physiology* **282**, C595–C605. (doi:10.1152/ajpcell.00270.2001)

39. del Alamo JC, Meili R, Alonso-Latorre B, Rodriguez-Rodriguez J, Aliseda A, Firtel RA, Lasheras JC. 2007 Spatio-temporal analysis of eukaryotic cell motility by improved force cytometry. *Proceedings of the National Academy of Sciences* **104**, 13343–13348. (doi:10.1073/pnas.0705815104)

40. Trepat X, Wasserman MR, Angelini TE, Millet E, Weitz DA, Butler JP, Fredberg JJ. 2009 Physical forces during collective cell migration. *Nature Physics* **5**, 426–430. (doi:10.1038/nphys1269)

41. Maruthamuthu V, Sabass B, Schwarz US, Gardel ML. 2011 Cell-ECM traction force modulates endogenous tension at cell-cell contacts. *Proceedings of the National Academy of Sciences* **108**, 4708–4713. (doi:10.1073/pnas.1011123108)

42. Rizvi SA, Neidt EM, Cui J, Feiger Z, Skau CT, Gardel ML, Kozmin SA, Kovar DR. 2009 Identification and Characterization of a Small Molecule Inhibitor of Formin-Mediated Actin Assembly. *Chemistry & Biology* **16**, 1158–1168. (doi:10.1016/j.chembiol.2009.10.006)

43. Isogai T, van der Kammen R, Innocenti M. 2015 SMIFH2 has effects on Formins and p53 that perturb the cell cytoskeleton. *Scientific Reports* **5**. (doi:10.1038/srep09802)

44. Acharya BR, Wu SK, Lieu ZZ, Parton RG, Grill SW, Bershadsky AD, Gomez GA, Yap AS. 2017 Mammalian Diaphanous 1 Mediates a Pathway for E-cadherin to Stabilize Epithelial Barriers through Junctional Contractility. *Cell Reports* **18**, 2854–2867. (doi:10.1016/j.celrep.2017.02.078)

45. Théry M, Pépin A, Dressaire E, Chen Y, Bornens M. 2006 Cell distribution of stress fibres in response to the geometry of the adhesive environment. *Cell Motil. Cytoskeleton* **63**, 341–355. (doi:10.1002/cm.20126)

46. Oakes PW, Banerjee S, Marchetti MC, Gardel ML. 2014 Geometry Regulates Traction Stresses in Adherent Cells. *Biophysical Journal* **107**, 825–833. (doi:10.1016/j.bpj.2014.06.045)

47. Sadati M, Taheri Qazvini N, Krishnan R, Park CY, Fredberg JJ. 2013 Collective migration and cell jamming. *Differentiation* **86**, 121–125. (doi:10.1016/j.diff.2013.02.005)

48. He L, Wang X, Tang HL, Montell DJ. 2010 Tissue elongation requires oscillating contractions of a basal actomyosin network. *Nature Cell Biology* **12**, 1133–1142. (doi:10.1038/ncb2124)

49. Chiou KK, Hufnagel L, Shraiman BI. 2012 Mechanical Stress Inference for Two Dimensional Cell Arrays. *PLoS Computational Biology* **8**, e1002512. (doi:10.1371/journal.pcbi.1002512)

50. Kasza KE, Farrell DL, Zallen JA. 2014 Spatiotemporal control of epithelial remodeling by regulated myosin phosphorylation. *Proceedings of the National Academy of Sciences* **111**, 11732–11737. (doi:10.1073/pnas.1400520111)

51. Brodland GW, Veldhuis JH, Kim S, Perrone M, Mashburn D, Hutson MS. 2014 CellFIT: A Cellular Force-Inference Toolkit Using Curvilinear Cell Boundaries. *PLoS ONE* **9**, e99116. (doi:10.1371/journal.pone.0099116)

52. Tetley RJ, Staddon MF, Banerjee S, Mao Y. 2018 Tissue Fluidity Promotes Epithelial Wound Healing. *bioRxiv* (doi:10.1101/433557)

53. Manning ML, Foty RA, Steinberg MS, Schoetz E-M. 2010 Coaction of intercellular adhesion and cortical tension specifies tissue surface tension. *Proceedings of the National Academy of Sciences* **107**, 12517–12522. (doi:10.1073/pnas.1003743107)

54. Koride S, Loza AJ, Sun SX. 2018 Epithelial vertex models with active biochemical regulation of contractility can explain organized collective cell motility. *APL Bioengineering* **2**, 031906. (doi:10.1063/1.5023410)

55. Bar-Kochba E, Toyjanova J, Andrews E, Kim K-S, Franck C. 2015 A Fast Iterative Digital Volume Correlation Algorithm for Large Deformations. *Experimental Mechanics* **55**, 261–274. (doi:10.1007/s11340-014-9874-2)

56. Treloar KK, Simpson MJ. 2013 Sensitivity of Edge Detection Methods for Quantifying Cell Migration Assays. *PLoS ONE* **8**, e67389. (doi:10.1371/journal.pone.0067389)

57. Mashburn DN, Lynch HE, Ma X, Hutson MS. 2012 Enabling user-guided segmentation and tracking of surface-labeled cells in time-lapse image sets of living tissues. *Cytometry Part A* **81A**, 409–418. (doi:10.1002/cyto.a.22034)

58. Püspöki Z, Storath M, Sage D, Unser M. 2016 Transforms and Operators for Directional Bioimage Analysis: A Survey. In *Focus on Bio-Image Informatics* (eds WH De Vos, S Munck, J-P Timmermans), pp. 69–93. Cham: Springer International Publishing. (doi:10.1007/978-3-319-28549-8_3)